\newif\ifCM 
\newif\ifSF 
\CMfalse
\SFfalse
\ifCM
\documentclass[envcountsame]{svjour3}
\else
\documentclass[11pt,a4paper]{article}
\fi
\ifCM
\usepackage[caption=false]{subfig}
\smartqed
\else
\usepackage{subfig}
\usepackage[ocgcolorlinks,allcolors={blue}]{hyperref}
\fi
\usepackage[utf8]{inputenc}
\usepackage[T1]{fontenc}
\usepackage{amsmath}
\usepackage{amssymb}
\usepackage{graphicx}
\usepackage[usenames,dvipsnames,svgnames,table]{xcolor}
\usepackage{multirow}
\usepackage{cite}
\usepackage{xspace}
\usepackage{ifthen}
\usepackage{ifmtarg}
\ifCM
\else
\fi
\ifSF
\else
\graphicspath{{Images/}}
\fi
\makeatletter

\DeclareFontFamily{U}{mathx}{\hyphenchar\font45}
\DeclareFontShape{U}{mathx}{m}{n}{
      <5> <6> <7> <8> <9> <10>
      <10.95> <12> <14.4> <17.28> <20.74> <24.88>
      mathx10
      }{}
\DeclareSymbolFont{mathx}{U}{mathx}{m}{n}
\DeclareFontSubstitution{U}{mathx}{m}{n}
\DeclareMathSymbol{\bigtimes}{1}{mathx}{"91}

\newcommand*{\Reel}{{\mathbb{R}}}      
\newcommand*{\Poly}{{\mathbb{P}}}      
\newcommand*{\No}{\vecteur{N}}          
\newenvironment{Proof}{\noindent \textbf{Proof}.}{
\ifCM
\qed
\else
$\square$
\fi
}
\newcommand*{\CA}{\texttt{code\_aster}} 

\newcommand*{\psv}[3][]{
	\@ifmtarg{#1}{
                (#2,#3) 
        }{%
                (#2,#3)_{\vecteur{L}^2(#1)}
        }%
} 
\newcommand*{\psm}[3][]{
	\@ifmtarg{#1}{
                (#2,#3) 
        }{%
                (#2,#3)_{\matrice{L}^2(#1)}
        }%
} 
\newcommand*{\ps}[3][]{
	\@ifmtarg{#1}{
                (#2,#3) 
        }{%
                (#2,#3)_{#1}
        }%
} 
\newcommand*{\normev}[2][]{
	\@ifmtarg{#1}{
                \Vert #2 \Vert 
        }{%
                \Vert #2 \Vert_{\vecteur{L}^2(#1)}
        }%
}
\newcommand*{\normem}[2][]{
	\@ifmtarg{#1}{
                \Vert #2 \Vert 
        }{%
                \Vert #2 \Vert_{\matrice{L}^2(#1)}
        }%
}
\newcommand*{\norme}[2][]{
	\@ifmtarg{#1}{
                \Vert #2 \Vert 
        }{%
                \Vert #2 \Vert_{#1}
        }%
}
\newcommand{\SCAL}{{\cdot}}
\newcommand*{\snorme}[2][]{
	\@ifmtarg{#1}{
                \vert #2 \vert 
        }{%
                \vert #2 \vert_{#1}
        }%
} 

\newcommand*{\abs}[1]{\vert #1 \vert} 

\newcommand*{\tenseur}[3][\boldsymbol]{%
    \ifthenelse{#2=0}{#3}{%
	\ifthenelse{#2=1}{%
						\underline{#1{#3}}
					}{
		\ifthenelse{#2=2}{%
							\underline{\underline{ #1{#3}}}
						}{
			\ifthenelse{#2=3}{%
								\underline{\underline{\underline{#1{#3}}}}
							}{
				\ifthenelse{#2=4}{%
				\underline{\underline{#1{\mathbb{#3}}}}
				}{ERROR}
			}
		}
	}
}}

\newcommand*{\vecteur}[1]{\tenseur{1}{#1}} 
\newcommand*{\matrice}[1]{\tenseur{2}{#1}} 
\newcommand*{\divergence}[1]{\nabla {\cdot} #1 } 
\newcommand{\gradX}{\matrice{\nabla}_X}
\newcommand{\PK}{\matrice{P}} 
\newcommand{\Fdef}{\matrice{F}} 
\newcommand{\loadext}{\vecteur{f}} 
\newcommand{\Tn}{\vecteur{t}} 
\newcommand{\Xo}{\vecteur{X}} 
\newcommand{\xc}{\vecteur{x}} 

\newcommand{\moduletangent}{{\tenseur{4}{\mathbb{A}}}} 
\newcommand*{\Bn}{\Gamma_n}
\newcommand*{\Bd}{\Gamma_d}
\newcommand{\dV}{\: d\Omega_0}
\newcommand{\dA}{\: d\Gamma}
\newcommand{\dT}{{\partial T}}
\newcommand{\FT}{\mathcal{F}_{\dT}} 
\newcommand{\Fh}{\mathcal{F}_h} 
\newcommand{\Th}{\mathcal{T}_h} 
\newcommand{\Fhi}{\mathcal{F}_h^i} 
\newcommand{\Fhb}{\mathcal{F}_h^b} 
\newcommand{\Fhbd}{\mathcal{F}_h^{b,d}} 
\newcommand{\Fhbn}{\mathcal{F}_h^{b,n}} 
\newcommand{\nTF}{\vecteur{n}{}_{TF}} 
\newcommand{\nT}{\vecteur{n}{}_{T}} 
\newcommand{\Pkd}{\Poly_d^k} 
\newcommand{\Pkpd}{\Poly_d^{k+1}} 
\newcommand{\RTNkd}{\mathbb{RTN}_d^k} 
\newcommand{\PkF}{\Poly_{d-1}^k} 
\newcommand{\UkT}{\vecteur{U}^k_T} 
\newcommand{\Ukhz}{\vecteur{U}^k_{h,0}} 
\newcommand{\Ukhd}{\vecteur{U}^k_{h,d}} 
\newcommand{\Ukh}{\vecteur{U}^k_h} 
\newcommand{\vT}{\vecteur{v}_T} 
\newcommand{\vF}{\vecteur{v}_F} 
\newcommand{\vdT}{\vecteur{v}_{\partial T}} 
\newcommand{\vTh}{\vecteur{v}_{\Th}} 
\newcommand{\vFh}{\vecteur{v}_{\Fh}} 
\newcommand{\uT}{\vecteur{u}_T} 
\newcommand{\udT}{\vecteur{u}_{\dT}} 
\newcommand{\uTh}{\vecteur{u}_{\Th}} 
\newcommand{\uFh}{\vecteur{u}_{\Fh}} 

\newcommand{\vh}{\vecteur{v}_h}

\newcommand{\duT}{ \delta\vecteur{u}_T}
\newcommand{\dvT}{\delta \vecteur{v}_T}
\newcommand{\dudT}{ \delta\vecteur{u}_{\dT}}
\newcommand{\dvdT}{\delta \vecteur{v}_{\dT}}
\newcommand{\dvF}{\delta \vecteur{v}_F}
\newcommand{\duTh}{ \delta\vecteur{u}_{\Th}}
\newcommand{\dvTh}{\delta \vecteur{v}_{\Th}}
\newcommand{\duFh}{ \delta\vecteur{u}_{\Fh}}
\newcommand{\dvFh}{\delta \vecteur{v}_{\Fh}}
\newcommand{\vu}{\vecteur{u}}
\newcommand{\vv}{\vecteur{v}}

\newcommand{\GrT}{\matrice{G}_T} 
\newcommand{\Grh}{\matrice{G}_h} 
\newcommand{\RecT}{\matrice{R}(T;\Reel^{d\times d})} 
\newcommand{\FrT}{\matrice{F}_T} 
\newcommand{\SdTk}{\vecteur{S}_{\dT}^k}
\newcommand{\tSdTk}{\vecteur{\tilde S}_{\dT}^k}
\newcommand{\hSdTk}{\vecteur{\hat S}_{\dT}^k}
\newcommand{\hSdTks}{\vecteur{\hat S}_{\dT}^{k*}}
\newcommand{\DTkp}{\vecteur{D}_T^{k+1}}
\newcommand{\PikT}{\vecteur{\Pi}^{k}_T} 
\newcommand{\PikF}{\vecteur{\Pi}^{k}_F} 
\ifCM
\else
\newtheorem{theorem}{Theorem}
\newtheorem{lemma}[theorem]{Lemma}

\newtheorem{remark}[theorem]{Remark}

\fi
\makeatother
\ifCM
\journalname{Computational Mechanics}
\fi
\begin{document}
\ifCM
\title{Hybrid High-Order methods for finite deformations of hyperelastic materials}
 \titlerunning{A Hybrid High-Order methods for finite deformations of hyperelastic materials}
\author{Mickaël Abbas \and Alexandre Ern \and Nicolas Pignet
}                     
\institute{Mickaël Abbas and Nicolas Pignet \at EDF R\&D,
              7 Boulevard Gaspard Monge, 91120 Palaiseau, France
              and IMSIA, UMR EDF/CNRS/CEA/ENSTA 9219, 828 Boulevard des Maréchaux,
91762 Palaiseau Cedex, France,
              \email{mickael.abbas@edf.fr, nicolas.pignet@edf.fr}  
           \and
                     Alexandre Ern and Nicolas Pignet  \at
              Université Paris-Est, CERMICS (ENPC), 6-8 avenue Blaise Pascal, 77455 Marne la Vall\'ee cedex 2, and INRIA Paris, 75589 Paris, France,
              \email{alexandre.ern@enpc.fr, nicolas.pignet@enpc.fr} 
}
%
\date{Received: date / Revised version: date}
\maketitle
\else
\author{Mickael Abbas$^1$, Alexandre Ern$^2$, Nicolas Pignet$^{1,2}$}
\title{Hybrid High-Order methods for finite deformations of hyperelastic materials}
\maketitle
\footnotetext[1]{EDF R\&D, 7 Boulevard Gaspard Monge, 91120 Palaiseau,
  France and IMSIA, UMR EDF/CNRS/CEA/ENSTA 9219, 828 Boulevard des Maréchaux,
91762 Palaiseau Cedex, France}
\footnotetext[2]{Université Paris-Est, CERMICS (ENPC), 6-8 avenue Blaise Pascal, 77455 Marne-la-Vall\'ee cedex 2, and INRIA Paris, 75589 Paris, France}
\fi
\ifCM
\begin{abstract}
We devise and evaluate numerically Hybrid High-Order (HHO) methods for 
hyperelastic materials undergoing finite deformations. The HHO methods use as discrete unknowns piecewise polynomials of order $k\ge1$ on the mesh skeleton, together with cell-based polynomials that can be eliminated locally by static condensation. The discrete problem is written as the minimization of a broken nonlinear elastic energy where a local reconstruction of the displacement gradient is used. Two HHO methods are considered: a stabilized method where the gradient is reconstructed as a tensor-valued polynomial of order $k$ and a stabilization is added to the discrete energy functional, and an unstabilized method which reconstructs a stable higher-order gradient and circumvents the need for stabilization. Both methods satisfy the principle of virtual work locally with equilibrated tractions. We present a numerical study of the two HHO methods on test cases with known solution and on more challenging three-dimensional test cases including finite deformations with strong shear layers and cavitating voids. We assess the computational efficiency of both methods, and we compare our results to those obtained with an industrial software using conforming finite elements and to results from the literature. The two HHO methods exhibit robust behavior in the quasi-incompressible regime.
\end{abstract}
\else
\begin{abstract}
We devise and evaluate numerically Hybrid High-Order (HHO) methods for 
hyperelastic materials undergoing finite deformations. The HHO methods use as discrete unknowns piecewise polynomials of order $k\ge1$ on the mesh skeleton, together with cell-based polynomials that can be eliminated locally by static condensation. The discrete problem is written as the minimization of a broken nonlinear elastic energy where a local reconstruction of the displacement gradient is used. Two HHO methods are considered: a stabilized method where the gradient is reconstructed as a tensor-valued polynomial of order $k$ and a stabilization is added to the discrete energy functional, and an unstabilized method which reconstructs a stable higher-order gradient and circumvents the need for stabilization. Both methods satisfy the principle of virtual work locally with equilibrated tractions. We present a numerical study of the two HHO methods on test cases with known solution and on more challenging three-dimensional test cases including finite deformations with strong shear layers and cavitating voids. We assess the computational efficiency of both methods, and we compare our results to those obtained with an industrial software using conforming finite elements and to results from the literature. The two HHO methods exhibit robust behavior in the quasi-incompressible regime.
\end{abstract}
-- Quasi-incompressible materials
\\
\fi


\section{Introduction}
Hybrid-High Order (HHO) methods have been introduced a couple of years
ago for linear elasticity problems in \cite{DiPEr:2015} and for
diffusion problems in \cite{DiPEL:2014}. A review on diffusion problems
can be found in \cite{DiPEL:2016}, and a P\'eclet-robust analysis for
advection-diffusion problems in \cite{DPDEr:15}. Moreover,
an open-source implementation of HHO methods using 
generic programming tools is available through the \texttt{Disk++}
library described in \cite{CicDE:2017Submitted}.
Recent developments of HHO methods in computational mechanics include the incompressible Stokes equations
(with possibly large irrotational forces) \cite{DPELS:16}, the
incompressible Navier--Stokes equations \cite{DiPKr:2016submitted}, 
Biot's consolidation problem \cite{BoBoD:2016}, and
nonlinear elasticity with small deformations
\cite{BoDPS:2017}. The goal of the present work is to devise
and evaluate numerically HHO methods for 
hyperelastic materials undergoing finite deformations. Such problems are particularly challenging since
finite deformations induce an additional geometric nonlinearity on top of the one
present in the stress-strain constitutive relation. Moreover, hyperelastic materials
are often considered near the incompressible limit, so that robustness
in this situation is important. 

The discrete unknowns in HHO methods are face-based unknowns that are
piecewise polynomials of some order $k\ge1$ on the mesh
skeleton ($k\ge0$ for diffusion equations). 
Cell-based unknowns are also introduced in the discrete formulation.
These additional unknowns are instrumental for the stability
and approximation properties of the method and can be locally eliminated by using the
well-known static condensation technique. In the present nonlinear
context, this elimination is performed at each step of the nonlinear
iterative solver (typically Newton's method). The devising
of HHO methods hinges on two ideas: \textup{(i)} a reconstruction
operator that reconstructs locally from the local cell and
face unknowns a displacement field or a
tensor-valued field representing its gradient; \textup{(ii)} a stabilization operator that enforces
in a weak sense on each mesh face the consistency between the local face
unknowns and the trace of the cell unknowns. 
A somewhat subtle design of the stabilization operator
has been proposed in \cite{DiPEr:2015,DiPEL:2014} leading to
$O(h^{k+1})$ energy-error estimates, where $h$ is the mesh-size, for 
linear diffusion and elasticity problems and smooth solutions. HHO
methods offer several advantages: \textup{(i)} the construction is
dimension-independent; \textup{(ii)} general meshes (including fairly
general polytopal mesh cells and non-matching interfaces) are supported;
\textup{(iii)} a local formulation using equilibrated fluxes is
available, and \textup{(iv)} HHO methods are computationally attractive
owing to the static condensation of the cell unknowns and the higher-order convergence rates. 

HHO methods have been bridged to 
Hybridizable Discontinuous Galerkin (HDG) methods in \cite{CoDPE:2016}.
HDG methods, as originally devised in \cite{CoGoL:09}, are formulated
in terms of a discrete triple which approximates the flux,
the primal unknown, and its trace on the mesh skeleton. The HDG
method is then specified by the discrete spaces for the above triple,
and the stabilization operator that enters the discrete equations
through the so-called numerical flux trace. The difference between HHO
and HDG methods is twofold: \textup{(i)} the HHO reconstruction operator
replaces the discrete HDG flux (a similar rewriting of an HDG method for
nonlinear elasticity can be found in \cite{KaLeC:2015}), and, more
importantly, \textup{(ii)} both HHO and HDG penalize in a least-squares
sense the difference between the discrete trace unknown and the trace of
the discrete primal unknown (with a possibly mesh-dependent weight), 
but HHO uses a non-local operator over each
mesh cell boundary that delivers one-order higher approximation than
just penalizing pointwise the difference as in HDG. 

Discretization methods for linear and nonlinear elasticity have
undergone a vigorous development over the last decade. For discontinuous
Galerkin (dG) methods, we mention in particular \cite{LeNSO:04,HanLa:02,CocSW:06} for linear
elasticity, and \cite{Eyck2006,Noels2006} for nonlinear elasticity. 
HDG methods for linear elasticity have been coined in \cite{Soon2009}
(see also \cite{CoGNPS:11} for incompressible Stokes flows),
and extensions to nonlinear elasticity can be found in
\cite{Soon2008,Nguyen2012a,KaLeC:2015}.
Other recent developments in the last few years include, among others, Gradient Schemes 
for nonlinear elasticity with small deformations \cite{Droniou2015},
the Virtual Element Method (VEM) for linear and nonlinear elasticity
with small \cite{BeiraodaVeiga2015} and finite deformations
\cite{Chi2017,Wriggers2017}, the (low-order) hybrid dG method with
conforming traces for nonlinear elasticity \cite{Wulfinghoff2017},
the hybridizable weakly conforming Galerkin method with nonconforming
traces for linear elasticity \cite{Kraemer2016}, the Weak Galerkin
method for linear elasticity \cite{WaWWZ:16}, and the discontinuous
Petrov--Galerkin method for linear elasticity \cite{CarHe:16}.

In the present work, we devise and evaluate numerically two HHO
methods to approximate hyperelastic
materials undergoing finite deformations. Following the ideas of \cite{Eyck2006,KaLeC:2015} developed
in the context of dG and HDG methods, both HHO discrete solutions are
formulated as stationary points of a discrete energy functional that is
defined from the exact energy functional by replacing the displacement
gradient in the 
Piola--Kirchhoff tensor by its reconstructed counterpart. In the first
HHO method, called stabilized HHO (sHHO), a quadratic term associated
with the HHO-stabilization operator is added to the discrete energy
functional. For linear elasticity, one recovers the original HHO method 
from \cite{DiPEr:2015} if the displacement gradient is reconstructed
locally in the tensor-valued polynomial space $\gradX \Pkpd(T;\Reel^d)$
where $k$ is the degree of the polynomials attached to the mesh skeleton and
$T$ is a generic mesh cell (and if the displacement divergence is
reconstructed in $\Pkd(T;\Reel)$); the notation is defined
more precisely in the following sections. In the present nonlinear
context, the gradient is reconstructed in $\Pkd(T;\Reel^{d\times d})$
(which is a strict superspace of $\gradX \Pkpd(T;\Reel^d)$); the same
reconstruction space is considered for HDG in \cite{KaLeC:2015} for nonlinear
elasticity with finite deformations (where the
stabilization operator is, however, different), and a similar choice
with symmetric-valued reconstructions is considered for HHO in
\cite{BoDPS:2017} for nonlinear elasticity with small
deformations. The main reason for reconstructing the gradient in a
larger space stems from the fact that the reconstructed gradient of a
test function acts against a discrete Piola--Kirchhoff tensor which is
not in gradient form. For a discussion and a numerical
example in the context of the Leray--Lions problem, we refer the reader
to \cite[\S4.1]{DiPietro2017Submitted}. 

In nonlinear elasticity, the use
of stabilization can lead to numerical difficulties since it is not
clear beforehand how large the stabilization parameter ought to be and
since a large value of this parameter can deteriorate the conditioning of the system and hamper the convergence of the iterative solvers; see 
\cite{Eyck2008,Eyck2008a} for a related discussion on dG methods and
\cite{BeiraodaVeiga2015,Chi2017} for VEM. Moreover, 
\cite[Section~4]{KaLeC:2015} presents an example where spurious
solutions can appear in an HDG discretization if the stabilization
parameter is not large enough. 
Motivated by these difficulties, we also consider a second
method called unstabilized HHO (uHHO). Inspired by the recent ideas in  
\cite{John2016} on stable dG methods without penalty parameters, we
consider an HHO method where the gradient is reconstructed in a
higher-order polynomial space, and no stabilization
is added to the discrete energy. Focusing for simplicity on matching simplicial
meshes, the reconstruction space can be (i)  
the Raviart--Thomas--N\'ed\'elec (RTN) space $\RTNkd(T;\Reel^{d\times d}) := 
\Pkd(T;\Reel^{d\times d}) \oplus (\Poly^{k,\textrm{H}}_d(T;\Reel^d) \otimes \vecteur{X})$, where $\Poly^{k,\textrm{H}}_d(T;\Reel^d)$ is the space composed of homogeneous polynomials of degree $k$, or (ii)
the (larger) polynomial space $\Pkpd(T;\Reel^{d\times d})$. 
For both choices, we prove, using the ideas
in~\cite{John2016}, that the reconstructed gradient is stable, thereby
circumventing the need to introduce and tune any stabilization parameter. 
Reconstructing the gradient in $\RTNkd(T;\Reel^{d\times d})$
leads to optimal $O(h^{k+1})$-convergence rates for linear problems 
and smooth solutions, 
Instead, reconstructing the gradient in $\Pkpd(T;\Reel^{d\times d})$
leads to $O(h^k)$-convergence rates for linear problems and
smooth solutions, i.e., the method still converges but at a suboptimal
order in ideal situations. The advantage of
reconstructing the gradient in $\Pkpd(T;\Reel^{d\times d})$ is,
however, that our numerical results indicate that the method is more robust
to handle strongly nonlinear problems. 

This paper is organized as follows. In Section~\ref{sec:model}, we
present the nonlinear hyperelasticity problem and we introduce
some basic notation. The two HHO methods are presented in
Section~\ref{sec:HHO}, where we also discuss some theoretical and
implementation aspects. Section~\ref{sec_numconv} then contains test
cases with analytical (or computable) solution. We first
consider three-dimensional traction test cases with
manufactured solution to assess the convergence rates delivered by sHHO
and uHHO in the nonlinear case. Then, we consider the dilatation of a
quasi-incompressible annulus; in this test case, proposed in
\cite[Section~5.2]{KaLeC:2015}, the exact solution can be approximated
to a very high accuracy by solving an ordinary 
differential equation in the radial
coordinate. We also compare the computational efficiency of both
methods, and we consider a continuous Galerkin (cG) approximation
based on $H^1$-conforming finite elements using the industrial software
\texttt{code\_aster} \cite{CodeAster}. Section~\ref{sec:3D} considers
three application-driven, three-dimensional examples: the indentation of
a compressible and quasi-incompressible rectangular block (where we also
provide a comparison with the industrial software
\texttt{code\_aster}), a hollow cylinder deforming under compression and
shear, and a sphere expanding 
under traction with two cavitating voids. These last
two examples are particularly challenging, and our results are compared
to the HDG solutions reported in \cite{KaLeC:2015}.

\section{The nonlinear hyperelasticity problem}\label{sec:model}
We are interested in finding the static equilibrium configuration of an
elastic continuum body that occupies the domain $\Omega_0$ in the
reference configuration and that undergoes finite deformations under the
action of a body force $\loadext$ in $\Omega_0$, a traction force $\Tn$
on the Neumann boundary $\Bn$, and a prescribed displacement $\vu{}_d$
on the Dirichlet boundary $\Bd$. Here, $\Omega_0 \subset \Reel^d$, $d
\in \{2,3\}$, is a bounded connected polytopal domain with unit outward
normal $\No$ and with Lipschitz boundary $\Gamma := \partial \Omega$
decomposed in the two relatively open subsets $\Bn$ and $\Bd$ such that
$\overline{\Bn} \cup \overline{\Bd} = \Gamma$, $\Bn \cap \Bd = \emptyset
$, and $\Bd$ has positive Hausdorff-measure (so as to prevent rigid-body
motions). In what follows, we write $v$ for scalar-valued fields, $\vv$
or $\vecteur{V}$ for vector-valued fields, $\matrice{V}$ for
second-order tensor-valued fields, and $\tenseur{4}{V}$ for fourth-order
tensor-valued fields.

As is customary for elasticity problems with finite deformations, we adopt the Lagrangian description (cf, e.g, the textbooks \cite{Bonet1997, Ciarlet1988}). Due to the deformation, a point $\Xo \in \Omega_0$ is mapped to a point $\xc = \Xo + \vu(\Xo)$ in the equilibrium configuration, where $\vu : \Omega_0 \rightarrow \Reel^d$ is the displacement mapping. The model problem consists in finding a displacement mapping $\vu : \Omega_0 \rightarrow \Reel^d$ satisfying the following equations:
\begin{subequations}\label{probleme_nonlineaire}
\begin{alignat}{2} 
     	-\mbox{\underline{Div}}_X(\PK) &= \loadext & \quad &\mbox{ in } \Omega_0, \label{eq_pnba} \\
    	\vu &= \vu{}_d & \quad  &\mbox{ on } \Bd, \label{eq_pnbb}\\
     	\PK \,  \No &= \Tn & \quad &\mbox{ on } \Bn,\label{eq_pnbc}
\end{alignat}
\end{subequations}
where $\PK := \PK(\Xo,\Fdef(\vu))$ is the first Piola--Kirchhoff stress tensor and $\Fdef(\vu) = \matrice{I} +\gradX \, \vu$ is the deformation gradient. The deformation gradient takes values in $\Reel^{d \times d}_{+}$ which is the set of $d \times d$ matrices with positive determinant. The governing equations~\eqref{probleme_nonlineaire} are stated in Lagrangian form; in particular, the gradient and divergence operators are taken with respect to the coordinate $\Xo$ of the reference configuration (we use the subscript $X$ to indicate it).

We restrict ourselves to bodies consisting of homogeneous hyperelastic materials for which there exists a strain energy density $\Psi(\Fdef)$ defined by a function $\Psi : \Reel^{d \times d}_{+} \rightarrow \Reel$. We assume that  the  first Piola--Kirchhoff stress tensor is defined as $\PK = \partial_{\Fdef} \Psi$ so that the associated elastic modulus is given by $\moduletangent = \partial_{\Fdef \, \Fdef}^2 \Psi$. We denote by $V$ the set of all kinematically admissible displacements which satisfy the Dirichlet condition~\eqref{eq_pnbb}, and we define the energy functional $\mathcal{E}: V \rightarrow \Reel$ such that
\begin{equation}\label{eq:def_calE}
\mathcal{E} (\vv) = \int_{\Omega_0} \Psi(\Fdef(\vv)) \dV - \int_{\Omega_0} \loadext\SCAL\vv \dV - \int_{\Bn} \Tn\SCAL\vv \dA.
\end{equation}
The static equilibrium problem~\eqref{probleme_nonlineaire} consists of seeking the stationary points of the energy functional $\mathcal{E}$ which satisfy the following weak form of the Euler--Lagrange equations:
\begin{equation}
0 =  D \mathcal{E} (\vu) [\delta \vv] = \int_{\Omega_0} \PK(\Fdef(\vu)) : \gradX(\delta \vv)  \dV - \int_{\Omega_0} \loadext\SCAL\delta \vv \dV - \int_{\Bn} \Tn\SCAL\delta \vv \dA,
\end{equation}
for all virtual displacements $\delta \vv$ satisfying a zero boundary condition on $\Bd$. We assume that the strain energy density function $\Psi$ is polyconvex (cf e.g.~\cite{Ball1976}) so that local minimizers of the energy functional exist. In the present work, we will mainly consider hyperelastic materials of Neohookean type extended to the compressible range such that
\begin{equation}\label{NeoLaw}
\Psi(\Fdef) = \frac{\mu}{2}  \left( \Fdef:\Fdef - d \right) - \mu \ln J + \frac{\lambda}{2} \Theta(J)^2,
\end{equation}
where $J\in\Reel_{>0}$ is the determinant of $\Fdef$, $\mu$ and $\lambda$ are material constants, and $\Theta : \Reel_{>0} \rightarrow \Reel$ is a smooth function such that $ \Theta(J) = 0 \Leftrightarrow J = 1$ and $\Theta'(1)  \neq 0$. The function $\Theta$ represents the volumetric deformation energy, and the potential $\Psi$ defined by~\eqref{NeoLaw} satisfies the principle of material frame indifference \cite{Ciarlet1988}. For further insight into the physical meaning, we refer the reader to \cite[Chap.7]{Ogden1997}. For later use, it is convenient to derive directly from \eqref{NeoLaw} the first Piola--Kirchhoff stress tensor
\begin{equation}
\PK(\Fdef) = \mu ( \Fdef - \Fdef^{-T}) +  \lambda J \Theta(J) \Theta'(J)  \Fdef^{-T},
\end{equation}
where we have used that $\partial_{\Fdef} J = J\Fdef^{-T}$, as well as the elastic modulus
\begin{align}
\moduletangent(\Fdef) =& \mu  ( \matrice{I}  \, \overline{\otimes} \, \matrice{I} + \Fdef^{-T} \underline{\otimes} \, \Fdef^{-1})  -\lambda  J \Theta(J) \Theta'(J)  \Fdef^{-T} \underline{\otimes} \, \Fdef^{-1} \nonumber \\
 & + \lambda  \left[ J\Theta(J)  (J \Theta''(J) + \Theta'(J)) + ( J \Theta'(J))^2 \right] \Fdef^{-T} \otimes \Fdef^{-T},
\end{align}
where $\otimes$, $\underline{\otimes}$ and $\overline{\otimes}$  are
defined such that $\{ \circ \otimes \bullet\}_{ijkl} = \{ \circ \}_{ij}
\{ \bullet\}_{kl}$, $\{ \circ \, \underline{\otimes} \, \bullet\}_{ijkl}
= \{ \circ \}_{il} \{ \bullet\}_{jk}$ and $\{ \circ \,
\overline{\otimes} \, \bullet\}_{ijkl} = \{ \circ \}_{ik} \{
\bullet\}_{jl}$, for all $1\le i,j,k,l\le d$.
\section{The Hybrid High-Order method}
\label{sec:HHO} 

In this section, we present the unstabilized and stabilized HHO methods to be
considered in our numerical tests. 

\subsection{Discrete setting}
Let $(\Th)_{h>0}$ be a shape-regular sequence of affine simplicial meshes with no hanging nodes of the domain $\Omega_0$. A generic mesh cell in $\Th$ is denoted $T\in\Th$, its diameter $h_T$, and its unit outward normal $\nT$. It is customary to define the global mesh-size as $h = \max_{T \in \Th} h_T$. The mesh faces are collected in the set $\Fh$, and a generic mesh face is denoted $F\in\Fh$. The set $\Fh$ is further partitioned into the subset $\Fhi$ which is the collection of mesh interfaces and the subset $\Fhb$ which is the collection of mesh faces located at the boundary $\Gamma$. We assume that the mesh is compatible with the partition of the boundary $\Gamma$ into $\Bd$ and $\Bn$, and we further split the set $\Fhb$ into the disjoint subsets $\Fhbd$ and $\Fhbn$ with obvious notation. 
For all $T \in \Th$, $\FT$ is the collection of the mesh faces that are subsets of $\partial T$.

\begin{figure}
    \centering
    \includegraphics[scale=0.25]{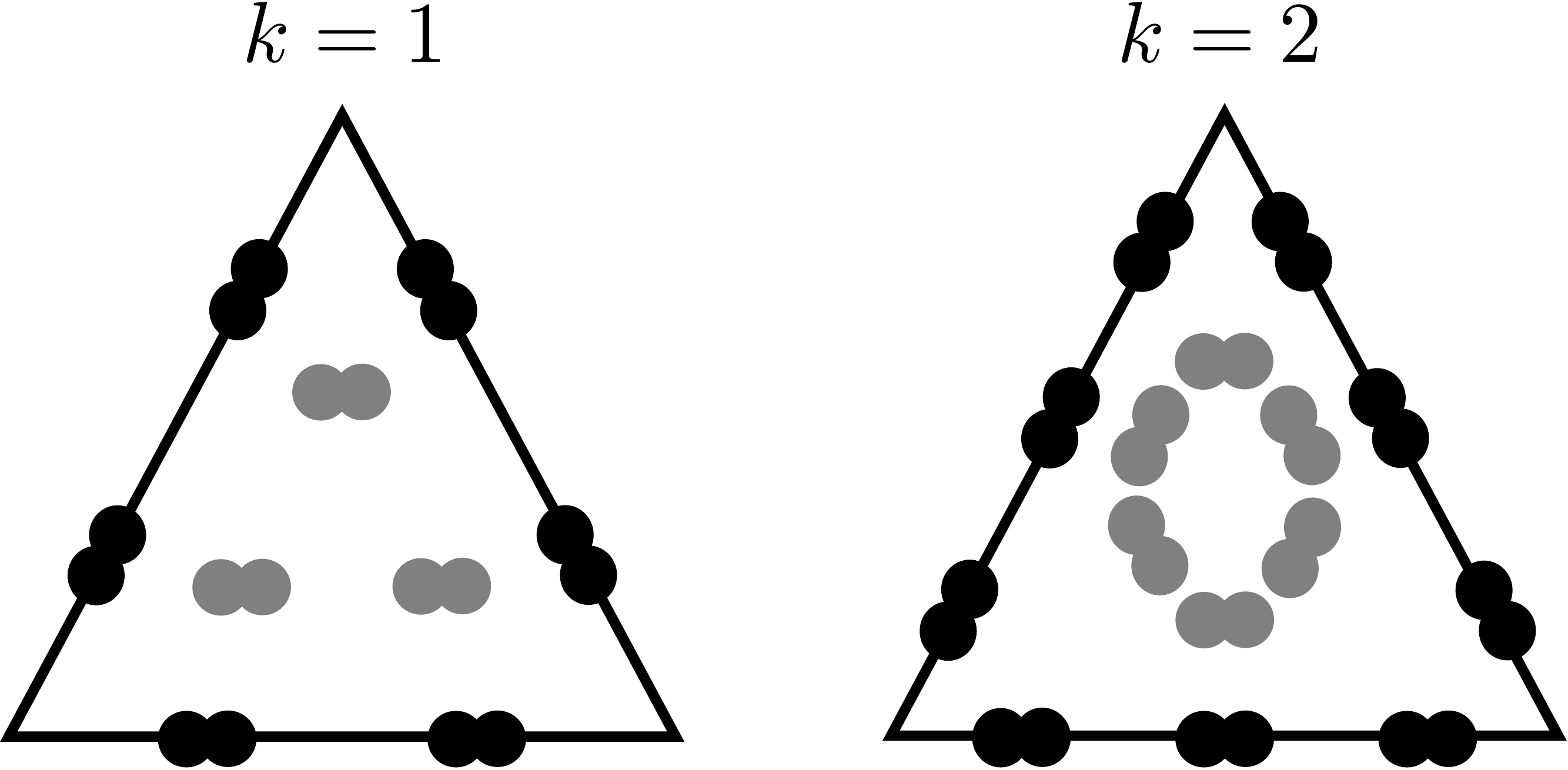} 
\caption{Face (black) and cell (gray) degrees of freedom in $\UkT$ for $k=1$ and $k=2$ in the two-dimensional case (each dot represents a degree of freedom which is not necessarily a point evaluation).}
\label{fig_HHO_dofs}
\end{figure}

Let $k \geq 1$ be a fixed polynomial degree. In each mesh cell $T\in\Th$, the local HHO unknowns are a pair $(\vT,\vdT)$, where the cell unknown 
$\vT\in \Pkd(T; \Reel^d)$ is a vector-valued $d$-variable 
polynomial of degree at most $k$ in the mesh cell $T$, and
$\vdT \in \PkF(\FT; \Reel^d) = \bigtimes_{F \in\FT} \PkF(F; \Reel^d)$ is a piecewise, vector-valued polynomial of degree at most $k$ on each face $F\in\FT$. 
We write more concisely that
\begin{equation}\label{local_dof}
(\vT,\vdT) \in \UkT := \Pkd(T;\Reel^d) \times \PkF(\FT;\Reel^d).
\end{equation}
The degrees of freedom are illustrated in Fig.~\ref{fig_HHO_dofs},
where a dot indicates one degree of freedom (and is not necessarily computed as a point evaluation). More generally, the polynomial degree $k$ of the face unknowns being fixed, HHO methods can be devised using cell unknowns that are polynomials of degree $l\in\{k-1,k,k+1\}$, see \cite{CoDPE:2016}; these variants are
not further considered herein.
We equip the space $\UkT$ with the following local discrete strain semi-norm:
\begin{equation} \label{eq:snorme}
\snorme[1,T]{( \vT, \vdT)}^2 := \normem[T]{\gradX \vT}^2 + \normev[\dT]{\gamma_{\dT}^{\frac12}(\vT-\vdT)}^2,
\end{equation}
with the piecewise constant function $\gamma_{\dT}$ such that $\gamma_{\dT|F}=h_F^{-1}$ for all $F\in\FT$ where $h_F$ is the diameter of $F$. We notice that 
$\snorme[1,T]{( \vT, \vdT)}=0$ implies that both functions $\vT$ and $\vdT$ are constant and take the same constant value.

\subsection{Local gradient reconstruction}
\label{sec:grad_rec}
A crucial ingredient in the devising of the HHO method is a local
gradient reconstruction in each mesh cell $T \in \Th$. This reconstruction
is materialized by an operator $\GrT : \UkT \rightarrow \RecT$, where
$\RecT$ is some finite-dimensional linear space typically composed of $\Reel^{d\times d}$-valued 
polynomials in $T$. For all $(\vT,\vdT) \in \UkT$, the reconstructed
gradient $\GrT(\vT,\vdT) \in \RecT$ is obtained by solving the 
following local problem: For all $\matrice{\tau} \in \RecT$,
\begin{equation}\label{eq_reconstruction_grad}
\psm[T]{\GrT( \vT,\vdT)}{\matrice{\tau}} = \psm[T]{\gradX{\vT}}{\matrice{\tau}} + \psv[\dT]{\vdT - \vT}{\matrice{\tau} \, \nT}.
\end{equation} 
Solving this problem entails inverting the mass matrix associated with some
basis of the polynomial space $\RecT$.
In the present work, we consider three choices for the reconstruction
space $\RecT$. 
The choice $\RecT := \Pkd(T;\Reel^{d\times d})$ is considered in the 
context of the stabilized HHO method which is further described in 
Section~\ref{sec:sHHO}. The other two choices are 
$\RecT = \RTNkd(T;\Reel^{d\times d})$ (that is, the 
RTN space of order $k$ defined in the introduction) and the larger space
$\RecT = \Pkpd(T;\Reel^{d\times d})$. 
These choices are considered
in the context of the unstabilized HHO method which is further described
in Section~\ref{sec:uHHO}. 

\begin{lemma}[Boundedness and stability]\label{lemma_bnd_stab}
The gradient reconstruction operator defined by~\eqref{eq_reconstruction_grad} enjoys the following properties: 
\textup{(i)} Boundedness: There is $\alpha_\sharp$, uniform w.r.t.~$h$, so that, for all $T\in\Th$,
\begin{equation} \label{eq:G_bnd}
\normem[T]{\GrT( \vT,\vdT)} \le \alpha_\sharp\snorme[1,T]{( \vT, \vdT)},
\qquad \forall (\vT,\vdT) \in \UkT.
\end{equation}
\textup{(ii)} Stability: Provided $\RTNkd(T;\Reel^{d\times d})\subseteq \RecT$,
there is $\alpha_\flat>0$, uniform w.r.t.~$h$, so that, for all $T\in\Th$,
\begin{equation} \label{eq:G_stab}
\normem[T]{\GrT( \vT,\vdT)} \ge \alpha_\flat\snorme[1,T]{( \vT, \vdT)},
\qquad \forall (\vT,\vdT) \in \UkT
\end{equation}
\end{lemma}

\begin{Proof}
The boundedness property~\eqref{eq:G_bnd} follows by applying the Cauchy--Schwarz inequality to the right-hand side of~\eqref{eq_reconstruction_grad} and a discrete trace inequality so as to bound $\normev[\dT]{\matrice{\tau} \, \nT}$ by $h_T^{-\frac12}\normem[T]{\matrice{\tau}}$.
The proof of the stability property~\eqref{eq:G_stab} is inspired from \cite{John2016}; we sketch it for completeness. Let $(\vT,\vdT) \in \UkT$. We need to
find a field $\matrice{\tau}\in\RecT$ so that 
\textup{(i)} 
$\snorme[1,T]{(\vT, \vdT)}^2 \le c\psm[T]{\GrT(\vT,\vdT)}{\matrice{\tau}}$
and 
\textup{(ii)} 
$\normem[T]{\matrice{\tau}} \le c\snorme[1,T]{(\vT,\vdT)}$ for some constant
$c$ uniform w.r.t.~$h$. Owing to our assumption
$\RTNkd(T;\Reel^{d\times d})\subseteq \RecT$, we can build 
$\matrice{\tau}\in \RTNkd(T;\Reel^{d\times d})$, and we do so by prescribing its
canonical degrees of freedom in $T$ as follows:
\begin{alignat*}{2} 
&\psm[T]{\matrice{\tau}}{\matrice{\phi}} = \psm[T]{\gradX \vT}{\matrice{\phi}}, &\quad& \forall\matrice{\phi} \in \Poly^{k-1}_d(T;\Reel^{d \times d}), \\
&\psv[\dT]{\matrice{\tau}\,\nT}{\vecteur{\varphi}} = \psv[\dT]{\gamma_{\dT}(\vdT - \vT)}{\vecteur{\varphi}}, &\quad& \forall \vecteur{\varphi} \in \PkF(\FT;\Reel^d). 
\end{alignat*}
With this choice, the above property~(i) holds true since
$\psm[T]{\GrT(\vT,\vdT)}{\matrice{\tau}}=\snorme[1,T]{(\vT, \vdT)}^2$,
whereas (ii) can be shown by using the classical stability of 
RTN functions in terms of their canonical degrees of freedom. 
\end{Proof}

\begin{remark}[General meshes]
The above stability proof exploits the properties of the RTN functions on simplicial meshes. If the meshes contain hanging nodes or cells with more general shapes, one possibility considered in the recent work \cite{DiPietro2017Submitted} is to reconstruct the gradient using piecewise RTN functions on a simplicial submesh of the mesh cell $T\in\Th$. Another construction has been recently devised in \cite{Eymard2017Preprint} for dG methods using a high-order lifting of the jumps on a simplicial submesh.
\end{remark}

\subsection{The unstabilized HHO method}
\label{sec:uHHO}

Let us set $\Pkd(\Th;\Reel^d) :=\bigtimes_{T \in\Th} \Pkd(T; \Reel^d)$
and $\PkF(\Fh;\Reel^d) := \bigtimes_{F \in\Fh} \PkF(F; \Reel^d)$.
The global space of discrete HHO unknowns is defined as
\begin{equation} \label{eq:def_Ukh}
\Ukh := \Pkd(\Th;\Reel^d) \times \PkF(\Fh;\Reel^d).
\end{equation}
For an element $\vh \in \Ukh$, we use the notation $\vh = (\vTh,\vFh)$.
For any mesh cell $T \in \Th$, we denote by $(\vT,\vdT)\in \UkT$ the local components of $\vh$ attached to the mesh cell $T$ and the faces composing its boundary, and for any mesh face $F\in\Fh$, we denote by $\vF$ the component attached to the face $F$. The Dirichlet boundary condition on the displacement field can be enforced explicitly on the discrete unknowns attached to the boundary faces in $\Fhbd$. We set
\begin{subequations}\begin{align}
\Ukhd &:= \left\lbrace (\vTh, \vFh ) \in \Ukh \: \vert \: \vF = \PikF(\vu_d), \; \forall F \in \Fhbd   \right \rbrace, \\
\Ukhz &:= \left\lbrace (\vTh, \vFh ) \in \Ukh \: \vert \: \vF = \vecteur{0}, \; \forall F \in \Fhbd   \right \rbrace,
\end{align}\end{subequations}
where $\PikF$ denotes the $L^2$-orthogonal projector onto $\PkF(F;\Reel^d)$.

The discrete counterpart of the energy functional $\mathcal{E}$ defined
by~\eqref{eq:def_calE} is the discrete energy functional $\mathcal{E}^{\textrm{u}}_h : \Ukh \rightarrow \Reel$ defined by
\begin{equation}\label{discete_energy}
\mathcal{E}^{\textrm{u}}_h (\vTh, \vFh )= \sum_{T\in\Th} \left\{
\int_T \Psi(\FrT (\vT,\vdT)) \:dT - \int_T \loadext\SCAL\vT \:dT\right\} - \sum_{F\in\Fhbn} \int_{F} \Tn\SCAL\vF \:dF,
\end{equation}
for all $(\vTh,\vFh) \in \Ukhd$, with the local deformation gradient operator $\FrT : \UkT \rightarrow \RecT$ such that $\FrT(\vT,\vdT) := \matrice{I} + \GrT( \vT,\vdT)$ where the local gradient reconstruction space is $\RecT = \RTNkd(T;\Reel^{d\times d})$ or $\RecT = \Pkpd(T;\Reel^{d\times d})$.

The discrete problem consists in seeking the stationary points of
the discrete energy functional $\mathcal{E}^{\textrm{u}}_h$. 
This leads to the following discrete equations: Find $(\uTh,\uFh)\in\Ukhd$ such that 
\begin{equation}\label{eq_points_critique}
\sum_{T\in\Th} \psm[T]{\PK(\FrT(\uT,\udT))}{\GrT(\dvT,\dvdT)} =
\sum_{T\in\Th}  \psv[T]{ \loadext}{\dvT} + \sum_{F\in\Fhbn} \psv[F]{\Tn}{\dvF},
\end{equation} 
for any generic virtual displacement $(\dvTh, \dvFh) \in \Ukhz$. The
discrete problem~\eqref{eq_points_critique} expresses the principle of
virtual work at the global level. As is often the case with discrete
formulations using face-based discrete unknowns, it is possible to
devise a local principle of virtual work in terms of face-based discrete
tractions that comply with the law of action and reaction. This has been
shown in \cite{CoDPE:2016} for HHO methods applied to the diffusion
equation, and the argument extends immediately to the present
context. Let $T\in\Th$ be a mesh cell and let $F\in\FT$ be one of its
faces. Let $\nTF$ denote the restriction to $F$ of the unit outward normal vector $\nT$. Let us define the discrete traction 
\begin{equation}\label{eq:traction}
\vecteur{T}_{T,F} = \PikF(\matrice{\Pi}_T^R(\PK(\FrT(\uT,\udT)))\SCAL\nTF),
\end{equation} 
where $\matrice{\Pi}_T^R$ denotes
the $L^2$-orthogonal projector onto $\RecT$. (Note that the projector $\PikF$ is not needed if $\RecT = \RTNkd(T;\Reel^{d\times d})$ since the normal component on $\dT$ of functions in $\RTNkd(T;\Reel^{d\times d})$ is in $\PkF(\dT;\Reel^d)$.)

\begin{lemma}[Equilibrated tractions] \label{lem:equil_uHHO}
The following local principle of virtual work holds true for all $T\in\Th$: For all $\dvT \in \Pkd(T;\Reel^d)$,
\begin{equation} \label{eq:pwk}
\psm[T]{\PK(\FrT(\uT,\udT))}{\gradX{\dvT}}-\sum_{F\in\FT}
\psv[F]{\vecteur{T}_{T,F}}{\dvT} = \psv[T]{ \loadext}{\dvT},
\end{equation}
where the discrete tractions $\vecteur{T}_{T,F}\in \PkF(F; \Reel^d)$ defined by~\eqref{eq:traction} satisfy the following law of action and reaction for all $F\in\Fhi\cup\Fhbn$:
\begin{subequations}\label{eq:balance}\begin{alignat}{2}
&\vecteur{T}_{T_-,F} + \vecteur{T}_{T_+,F} = \vecteur{0},
&\quad&\text{if $F\in\Fhi$ with $\partial T_- \cap \partial T_+ = F$},\\
&\vecteur{T}_{T,F}  = \PikF(\Tn),&\quad&\text{if $F\in\Fhbn$ with $\partial T\cap \Bn=F$}.
\end{alignat}
\end{subequations} 
\end{lemma}

\begin{Proof}
We follow the ideas in \cite{CoDPE:2016}. The local principle of virtual work~\eqref{eq:pwk} follows by considering the virtual displacement $((\dvT \delta_{T,T'})_{T'\in\Th},(\vecteur{0})_{F\in\Fh})\in\Ukhz$ in~\eqref{eq_points_critique}, with the Kronecker delta such that $\delta_{T,T'}=1$ if $T=T'$ and $\delta_{T,T'}=0$ otherwise, and observing that, owing to~\eqref{eq_reconstruction_grad}, we have
\begin{align*}
\psv[T]{ \loadext}{\dvT} ={}& 
\psm[T]{\PK(\FrT(\uT,\udT))}{\GrT(\dvT,\vecteur{0})} \\
= {}& \psm[T]{\matrice{\Pi}_T^R\PK(\FrT(\uT,\udT))}{\GrT(\dvT,\vecteur{0})} \\
= {}& \psm[T]{\matrice{\Pi}_T^R(\PK(\FrT(\uT,\udT)))}{\gradX{\dvT}} \\
&- \sum_{F\in\FT} \psv[F]{\matrice{\Pi}_T^R(\PK(\FrT(\uT,\udT))) \, \nTF}{\dvT}\\
={}&\psm[T]{\PK(\FrT(\uT,\udT))}{\gradX{\dvT}}-\sum_{F\in\FT}
\psv[F]{\vecteur{T}_{T,F}}{\dvT}.
\end{align*}
Similarly, the balance property~\eqref{eq:balance} follows by 
considering, for all $F\in\Fhi\cup\Fhbn$, the virtual displacement $((\vecteur{0})_{T\in\Th},(\dvF \delta_{F,F'})_{F'\in\Fh})\in\Ukhz$ in~\eqref{eq_points_critique} (with obvious notation for the face-based Kronecker delta), and observing that both $\dvF$ and $\vecteur{T}_{T_{\pm},F}$ are in $\PkF(F;\Reel^d)$.
\end{Proof}

Let us now discuss the choice of the gradient reconstruction space where one can set either $\RecT = \RTNkd(T;\Reel^{d\times d})$ or $\RecT = \Pkpd(T;\Reel^{d\times d})$. The key property with $\RecT = \RTNkd(T;\Reel^{d\times d})$ is that the normal component on $\dT$ of functions in $\RTNkd(T;\Reel^{d\times d})$ is in the space $\PkF(\dT;\Reel^d)$ used for the face HHO unknowns (the normal components of such functions actually span $\PkF(\dT;\Reel^d)$). Proceeding as in \cite{DiPEr:2015} then leads to the following important commuting property:
\begin{equation}\label{eq:commuting}
\GrT(\vecteur{I}_{T,\dT}(\vecteur{v})) = \matrice{\Pi}^R_T(\gradX\vecteur{v}),\qquad \forall \vecteur{v}\in H^1(T;\Reel^d),
\end{equation} 
where the reduction operator $\vecteur{I}_{T,\dT} : H^1(T;\Reel^d)\rightarrow \UkT$ is defined so that $\vecteur{I}_{T,\dT}(\vecteur{v})=(\vecteur{\Pi}^k_T(\vecteur{v}),\vecteur{\Pi}^k_{\dT}(\vecteur{v}))$, where $\vecteur{\Pi}^k_T$ is the $L^2$-orthogonal projector onto $\Pkd(T;\Reel^d)$ and $\vecteur{\Pi}^k_{\dT}$ is the $L^2$-orthogonal projector onto $\PkF(\FT;\Reel^d)$. Proceeding as in \cite[Thm.~8]{DiPEr:2015} and using the approximation properties of the RTN finite elements, one can show that for the linear elasticity problem and smooth solutions, the energy error measured as $\normem[\Th]{\gradX \vu - \Grh(\uTh, \uFh)}$ converges as $h^{k+1}|\vecteur{u}|_{\vecteur{H}^{k+2}(\Omega_0)}$ (the subscript $\matrice{L}^2(\Th)$ means that the Hilbertian sum of $L^2(T;\Reel^{d\times d})$-norms over the mesh cells is considered). Concerning implementation, we observe that the reconstruction operator needs to select basis functions for the RTN space; however, the canonical basis functions are not needed, and one can use simple monomial bases. 

Considering instead the choice $\RecT = \Pkpd(T;\Reel^{d\times d})$ leads to a larger space for the local gradient reconstruction (for $d=3$, the local space is of dimension $45$ ($k=1$) and $108$ ($k=2$) for RTN functions and of dimension $90$ ($k=1$) and $180$ ($k=2$) for $\Reel^{d\times d}$-valued polynomials of order $(k+1)$). One benefit of considering a larger space is, according to our numerical experiments, an increased robustness of the method to handle strongly nonlinear cases. One disadvantage is that the above property on the normal component of functions in $\RecT$ no longer holds. Therefore, one no longer has~\eqref{eq:commuting}; however, one can infer from~\eqref{eq_reconstruction_grad} the weaker property 
\begin{equation}\label{eq:commuting_tilde}
\GrT(\vecteur{\tilde I}_{T,\dT}(\vecteur{v})) = \gradX(\vecteur{\Pi}_T^k(\vecteur{v})),\qquad \forall \vecteur{v}\in H^1(T;\Reel^d),
\end{equation} 
where the reduction operator $\vecteur{\tilde I}_{T,\dT} : H^1(T;\Reel^d)\rightarrow \UkT$ is defined so that $\vecteur{\tilde I}_{T,\dT}(\vecteur{v})=(\vecteur{\Pi}^k_T(\vecteur{v}),\vecteur{\Pi}^k_{T}(\vecteur{v})_{|\dT})$. Proceeding as in \cite[Thm.~8]{DiPEr:2015}, one can show that for the linear elasticity problem and smooth solutions, the energy error $\normem[\Th]{\gradX \vu - \Grh(\uTh, \uFh)}$ converges as $h^{k}|\vecteur{u}|_{\vecteur{H}^{k+1}(\Omega_0)}$. This convergence rate will be confirmed by the experiments reported in Section~\ref{sec:manufactured}. Finally, regardless of the choice of $\RecT$, testing~\eqref{eq_reconstruction_grad} with a function $\matrice{\tau}=q\matrice{I} \in \Pkd(T;\Reel^{d\times d})$ with $q$ arbitrary in $\Pkd(T;\Reel)$, one can show that 
\begin{equation} \label{eq:trace}
\Pi_T^k(\mathop{\mathrm{tr}}(\GrT(\vecteur{I}_{T,\dT}(\vecteur{v}))) = \Pi_T^k(\divergence{\vecteur{v}}), \qquad \forall \vecteur{v}\in H^1(T;\Reel^d).
\end{equation}
The presence of the projector $\Pi_T^k$ on the left-hand side indicates
that $\mathop{\mathrm{tr}}(\GrT(\vecteur{I}_{T,\dT}(\vecteur{v})))$ may
be affected by a high-order perturbation hampering the argument of \cite[Prop.~3]{DiPEr:2015} to prove robustness in the quasi-incompressible limit for linear elasticity. Nevertheless, we observe absence of locking in the numerical experiments performed in Sections~\ref{sec:annulus} and~\ref{sec:3D}.

\subsection{The stabilized HHO method}
\label{sec:sHHO}

The discrete unknowns in the stabilized HHO method are exactly the same as those in the unstabilized HHO method. The only difference is in the form of the discrete elastic energy. In the stabilized HHO method, the gradient is reconstructed locally in the polynomial space $\RecT=\Pkd(T;\Reel^{d\times d})$ for all $T\in\Th$. Since the norm $\normem[T]{\GrT(\vT,\vdT)}$ does not control the semi-norm $\snorme[T]{(\vT,\vdT)}$ for all $(\vT,\vdT)\in \UkT$ (as can be seen from a simple counting argument based on the dimension of the involved spaces), we need to augment the discrete elastic energy by a stabilization semi-norm. This semi-norm is based on the usual stabilization operator for HHO methods $\SdTk : \UkT \rightarrow \PkF(\FT;\Reel^d)$ such that, for all $(\vT,\vdT)\in \UkT$,
\begin{equation} \label{eq:stab}
\SdTk(\vT,\vdT) = \vecteur{\Pi}_{\dT}^k\big(\vdT-\DTkp(\vT,\vdT)_{|\dT}-(\vT-\PikT(\DTkp(\vT,\vdT)))_{|\dT}\big),
\end{equation}
with the local displacement reconstruction operator $\DTkp : \UkT \rightarrow \Poly_d^{k+1} (T; \Reel^d)$ such that, for all $(\vT,\vdT) \in \UkT$, $\DTkp(\vT,\vdT) \in \Poly_d^{k+1}(T;\Reel^d)$ is obtained by solving the following Neumann problem in $T$: For all $\vecteur{w} \in \Poly_d^{k+1} (T;\Reel^d)$,
\begin{equation}\label{eq_reconstruction_depl}
\psm[T]{\gradX{\DTkp(\vT,\vdT)}}{\gradX{\vecteur{w}}} = \psm[T]{\gradX{\vT}}{\gradX{\vecteur{w}}} + \psv[\dT]{\vdT - \vT}{\gradX{\vecteur{w}} \, \nT},
\end{equation}
and additionally enforcing that $\int_T \DTkp(\vT,\vdT) \:dT= \int \vT \:dT$.
Comparing with~\eqref{eq_reconstruction_grad}, one readily sees that 
$\gradX{\DTkp(\vT,\vdT)}$ is the $L^2$-orthogonal projection of $\GrT(\vT,\vdT)$ onto the subspace $\gradX \Poly^{k+1}_d(T;\Reel^d) \subsetneq \Pkd(T;\Reel^{d \times d})= \RecT$. Following \cite[Lemma 4]{DiPEr:2015}, it is straightforward to establish the following stability and boundedness properties (the proof is omitted for brevity).

\begin{lemma}[Boundedness and stability]\label{lemma_stability_stab}
Let the gradient reconstruction operator be defined by~\eqref{eq_reconstruction_grad} with $\RecT=\Pkd(T;\Reel^{d\times d})$. Let the stabilization operator be defined by~\eqref{eq:stab}. Then, there exist real numbers $0<\alpha_\flat<\alpha_\sharp$, uniform w.r.t.~$h$, so that
\begin{equation}
\alpha_\flat \snorme[{1,T}]{(\vT, \vdT)} \leq \bigg(\normem[T]{\GrT (\vT, \vdT)}^2 + \normev[\dT]{\gamma_{\dT}^{\frac12}\SdTk(\vT,\vdT)}^2\bigg)^{\frac12} \leq \alpha_\sharp \snorme[{1,T}]{(\vT, \vdT)},
\end{equation}
for all $T \in \Th$ and all $(\vT,\vdT) \in \UkT$, with $\gamma_{\dT}$ defined below~\eqref{eq:snorme}.
\end{lemma}

\begin{remark}[HDG-type stabilization]
In general, HDG methods use the stabilization operator 
$\tSdTk(\vT,\vdT) = \vdT-\vT$ in the equal-order case, or
$\tSdTk(\vT,\vdT) = \vecteur{\Pi}_{\dT}^k(\vdT-\vT)$ if the cell unknowns are taken to be polynomials of order $(k+1)$ (see \cite{Lehrenfeld:10}). 
The definition in Eq.~\eqref{eq:stab}, introduced in \cite{DiPEr:2015}, enjoys, even in the equal-order case, the
high-order approximation property $\normev[\dT]{\gamma_{\dT}^{\frac12}\SdTk(\vecteur{I}_{T,\dT}(\vecteur{v}))} \le ch_T^{k+1}|\vecteur{v}|_{\vecteur{H}^{k+2}(T)}$ with the reduction operator $\vecteur{I}_{T,\dT} : H^1(T;\Reel^d)\rightarrow \UkT$ defined below~\eqref{eq:commuting} and $c$ uniform w.r.t.~$h$.
\end{remark}

In the stabilized HHO method, the discrete energy functional 
$\mathcal{E}^{\textrm{s}}_h : \Ukh \rightarrow \Reel$ is defined as
\begin{align}\label{discete_energy_stab}
\mathcal{E}^{\textrm{s}}_h (\vTh, \vFh) ={}& \sum_{T\in\Th} \left\{ \int_T \Psi(\FrT (\vT,\vdT)) - \int_T \loadext \SCAL \vT\: dT\right\} - \sum_{F\in\Fhbn} \int_F \Tn\SCAL \vF \:dF \nonumber \\
&+  \sum_{T\in\Th} \frac{\beta}{2} \psv[\dT]{\gamma_{\dT}\SdTk(\vT,\vdT)}{\SdTk(\vT,\vdT)},
\end{align}
with a user-dependent weight of the form $\beta=\beta_0\mu$ with typically $\beta_0\ge1$ (in the original HHO method for linear elasticity \cite{DiPEr:2015}, the choice $\beta_0=2$ is considered).
The discrete problem consists in seeking the stationary points of the discrete energy functional: 
Find $(\uTh,\uFh)\in\Ukhd$ such that 
\begin{align}\label{eq_points_critique_stab}
& \sum_{T\in\Th} \psm[T]{\PK(\FrT(\uT,\udT))}{\GrT(\dvT,\dvdT)} +  \sum_{T\in\Th} \beta \psv[\dT]{\gamma_{\dT}\SdTk(\uT,\udT)}{\SdTk(\dvT,\dvdT)} \nonumber  \\
 &= \sum_{T\in\Th}  \psv[T]{ \loadext}{\dvT} + \sum_{F\in\Fhbn} \psv[F]{\Tn}{\dvF},
\end{align}
for all $(\dvTh, \dvFh) \in \Ukhz$. As for the unstabilized HHO method, the discrete problem~\eqref{eq_points_critique_stab} expresses the principle of virtual work at the global level, and following~\cite{CoDPE:2016}, it is possible to devise a local principle of virtual work in terms of face-based discrete tractions that comply with the law of action and reaction. Let $T\in\Th$ be a mesh cell and let $F\in\FT$ be one of its faces. Let $\nTF$ denote the restriction to $F$ of the unit outward normal vector $\nT$. Let $\hSdTk : \PkF(\FT;\Reel^d) \rightarrow \PkF(\FT;\Reel^d)$ be defined such that 
\begin{equation} \label{eq:stab_hat}
\hSdTk(\vecteur{\theta}) = \vecteur{\Pi}_{\dT}^k\big(\vecteur{\theta}-(\matrice{I}-\PikT)\DTkp(\vecteur{0},\vecteur{\theta})\big).
\end{equation}
Comparing \eqref{eq:stab} with \eqref{eq:stab_hat}, we observe that $\SdTk(\vT,\vdT)=\hSdTk(\vdT-\vT)$ for all $(\vT,\vdT)\in\UkT$. Let $\hSdTks : \PkF(\FT;\Reel^d) \rightarrow \PkF(\FT;\Reel^d)$ be the adjoint operator of $\SdTk$ with respect to the $L^2(\dT;\Reel^d)$-inner product. We observe that the stabilization-related term in \eqref{eq_points_critique_stab} can be rewritten as
\begin{equation}
\psv[\dT]{\gamma_{\dT}\SdTk(\uT,\udT)}{\SdTk(\dvT,\dvdT)} =
\psv[\dT]{\hSdTks(\gamma_{\dT}\hSdTk(\udT-\uT))}{\dvdT-\dvT}.
\end{equation}
Finally, let us define the discrete traction 
\begin{equation}\label{eq:traction_stab}
\vecteur{T}_{T,F} = \matrice{\Pi}_T^k(\PK(\FrT(\uT,\udT)))\SCAL\nTF
+ \beta \hSdTks(\gamma_{\dT}\hSdTk(\udT-\uT)).
\end{equation} 

\begin{lemma}[Equilibrated tractions] \label{lem:equil_sHHO}
The following local principle of virtual work holds true for all $T\in\Th$: For all $\dvT \in \Pkd(T;\Reel^d)$,
\begin{equation} \label{eq:pwk_stab}
\psm[T]{\PK(\FrT(\uT,\udT))}{\gradX{\dvT}}-\sum_{F\in\FT}
\psv[F]{\vecteur{T}_{T,F}}{\dvT} = \psv[T]{ \loadext}{\dvT},
\end{equation}
where the discrete tractions $\vecteur{T}_{T,F} \in \PkF(F; \Reel^d)$ defined by~\eqref{eq:traction_stab} satisfy the following law of action and reaction for all $F\in\Fhi\cup\Fhbn$:
\begin{subequations}\label{eq:balance_stab}\begin{alignat}{2}
&\vecteur{T}_{T_-,F} + \vecteur{T}_{T_+,F} = \vecteur{0},
&\quad&\text{if $F\in\Fhi$ with $\partial T_- \cap \partial T_+ = F$},\\
&\vecteur{T}_{T,F}  = \PikF(\Tn),&\quad&\text{if $F\in\Fhbn$ with $\partial T\cap \Bn=F$}.
\end{alignat}
\end{subequations}  
\end{lemma}

\begin{Proof}
Proceed as in the proof of Lemma~\ref{lem:equil_uHHO}; see also \cite{CoDPE:2016}.
\end{Proof}

Let us briefly comment on the commuting properties of the reconstructed gradient in $\Pkd(T;\Reel^{d\times d})$. Proceeding as above, one obtains 
\begin{equation}\label{eq:commuting_stab}
\GrT(\vecteur{I}_{T,\dT}(\vecteur{v})) = \matrice{\Pi}^k_T(\gradX\vecteur{v}),\qquad \forall \vecteur{v}\in H^1(T;\Reel^d),
\end{equation}  
where the reduction operator $\vecteur{I}_{T,\dT} : H^1(T;\Reel^d)\rightarrow \UkT$ is defined below~\eqref{eq:commuting}. Proceeding as in \cite[Thm.~8]{DiPEr:2015}, one can show that for the linear elasticity problem and smooth solutions, the energy error $\normem[\Th]{\gradX \vu - \Grh(\uTh, \uFh)}$ converges as $h^{k+1}|\vecteur{u}|_{\vecteur{H}^{k+2}(\Omega_0)}$. This convergence rate will be confirmed by the experiments reported in Section~\ref{sec:manufactured}. Moreover, taking the trace in~\eqref{eq:commuting_stab}, we infer that (compare with~\eqref{eq:trace})
\begin{equation} \label{eq:trace_stab}
\mathop{\mathrm{tr}}(\GrT(\vecteur{I}_{T,\dT}(\vecteur{v})) = \Pi_T^k(\divergence{\vecteur{v}}), \qquad \forall \vecteur{v}\in H^1(T;\Reel^d),
\end{equation}
which is the key commuting property used in \cite{DiPEr:2015} to prove robustness for quasi-incompressible linear elasticity. This absence of locking is confirmed in the numerical experiments performed in Sections~\ref{sec:annulus} and~\ref{sec:3D} in the nonlinear regime. Finally, we refer the reader to \cite{BoDPS:2017} for further analytical results on symmetric-valued gradients reconstructed in the smaller space $\Pkd(T;\Reel_{\textrm{sym}}^{d\times d})$.
\begin{remark}[Choice of $\beta_0$]
For the HHO method applied to linear elasticity, a natural choice for the stabilization parameter is $\beta_0=2$ \cite{DiPEr:2015}. To our knowledge, there is no general theory on the choice of $\beta_0$ in the case of finite deformations of hyperelastic materials. Following ideas developed in \cite{Eyck2008, Eyck2008a} for dG and in \cite{BeiraodaVeiga2015} for VEM, one can consider to take (possibly in an adaptive fashion) the largest eigenvalue (in absolute value) of the elastic modulus $\moduletangent$. This choice introduces additional nonlinearities to be handled by Newton's method, and may require some relaxation. 
Another possibility discussed in \cite{Chi2017} for VEM methods is based
on the trace of the Hessian of the isochoric part of the strain-energy
density $\Psi$. Such an approach bears similarities with the classic
selective integration for FEM, and for the Neohookean materials
considered herein, this choice implies to take $\beta_0 = 1$. Finally,
let us mention that \cite[Section~4]{KaLeC:2015} presents an example
where spurious solutions can appear if the HDG stabilization parameter is
not large enough; however, too large values of the parameter can also deteriorate the conditioning number of the stiffness matrix and can cause numerical instabilities in Newton's method.
\end{remark}
\subsection{Nonlinear solver and static condensation}
\label{sec:implement}
Both nonlinear problems \eqref{eq_points_critique} and
\eqref{eq_points_critique_stab} are solved using Newton's method. Let
$n\ge0$ be the index of the Newton's step. Given an initial discrete
displacement $(\uTh, \uFh)^0\in\Ukhd$, one computes at each Newton's
step the incremental displacement $(\duTh, \duFh)^n\in \Ukhz$ and
updates the discrete displacement as $(\uTh, \uFh)^{n+1} = (\uTh,
\uFh)^n+(\duTh, \duFh)^n$. The linear system of equations to be solved is
\ifCM
\begin{align}\label{eq_stiffness_matrix}
& \hphantom{+} \sum_{T\in\Th} \psm[T]{\moduletangent(\FrT ( \uT, \udT)^n): \GrT(\duT,\dudT)^n}{\GrT(\dvT,\dvdT)} \nonumber \\ 
&+  \sum_{T\in\Th} \beta \psv[\dT]{\gamma_{\dT}\SdTk(\duT,\dudT)^n}{\SdTk(\dvT,\dvdT)} \nonumber \\
&= -R_h((\uTh, \uFh)^n, (\dvTh,\dvFh)),
\end{align}
\else
\begin{align}\label{eq_stiffness_matrix}
&\sum_{T\in\Th} \psm[T]{\moduletangent(\FrT ( \uT, \udT)^n): \GrT(\duT,\dudT)^n}{\GrT(\dvT,\dvdT)} \nonumber \\ 
&+  \sum_{T\in\Th} \beta \psv[\dT]{\gamma_{\dT}\SdTk(\duT,\dudT)^n}{\SdTk(\dvT,\dvdT)}= -R_h((\uTh, \uFh)^n, (\dvTh,\dvFh)),
\end{align}
\fi
for all $(\dvT,\dvdT)\in\Ukhz$, with the residual term
\ifCM
\begin{align}
& \hphantom{+} R_h((\uTh, \uFh)^n, (\dvTh, \dvFh)) \nonumber  \\ 
& = \sum_{T\in\Th} \psm[T]{\PK(\FrT(\uT,\udT)^n)}{\GrT(\dvT,\dvdT)} \nonumber \\
& + \sum_{T\in\Th} \beta \psv[\dT]{\gamma_{\dT}\SdTk(\uT,\udT)^n}{\SdTk(\dvT,\dvdT)} \nonumber \\ 
&- \sum_{T\in\Th}  \psv[T]{ \loadext}{\dvT} - \sum_{F\in\Fhbn} \psv[F]{\Tn}{\dvF},
\end{align}
\else
\begin{align}
& \hphantom{+} R_h((\uTh, \uFh)^n, (\dvTh, \dvFh)) \nonumber  \\ 
& = \sum_{T\in\Th} \psm[T]{\PK(\FrT(\uT,\udT)^n)}{\GrT(\dvT,\dvdT)} \nonumber \\
&+  \sum_{T\in\Th} \beta \psv[\dT]{\gamma_{\dT}\SdTk(\uT,\udT)^n}{\SdTk(\dvT,\dvdT)} \\ 
&- \sum_{T\in\Th}  \psv[T]{ \loadext}{\dvT} - \sum_{F\in\Fhbn} \psv[F]{\Tn}{\dvF} \nonumber,
\end{align}
\fi
where $\beta=0$ in the unstabilized case and $\beta=\beta_0\mu$ in the stabilized case, the gradient being reconstructed in the corresponding polynomial space.
It can be seen from~\eqref{eq_stiffness_matrix} that the assembling of the stiffness matrix on the left-hand side is local (and thus fully parallelizable).

As is classical with HHO methods \cite{DiPEr:2015}, and more generally with hybrid approximation methods, the cell unknowns $\duTh^n$ in~\eqref{eq_stiffness_matrix} can be eliminated locally using a static condensation (or Schur complement) technique. Indeed, testing~\eqref{eq_stiffness_matrix} against the function $((\dvT\delta_{T,T'})_{T'\in\Th},(\vecteur{0})_{F\in\Fh})$ with Kronecker delta $\delta_{T,T'}$ and $\dvT$ arbitrary in $\Pkd(T;\Reel^d)$, one can express, for all $T\in\Th$, the cell unknown $\duT^n$ in terms of the local face unknowns collected in $\dudT^n$. As a result, the static condensation technique allows one to reduce~\eqref{eq_stiffness_matrix} to a linear system in terms of the face unknowns only. This reduced system is of size $N_{\Fh} \times  \dim(\PkF(T; \Reel^d))$ where $N_{\Fh}$ denotes the number of mesh faces, 
and its stencil is such that each mesh face is connected to its neighbouring faces that share a mesh cell with the face in question. 

The implementation of the HHO methods is realized using the open-source library \texttt{DiSk++} \cite{CicDE:2017Submitted} which provides generic programming tools for the implementation of HHO methods and  is available at the address \texttt{https://github.com/wareHHOuse/diskpp}. The data structure requires access to faces and cells as in standard dG or HDG codes. The gradient and stabilization operators are built locally at the cell level using scaled translated monomials to define the basis functions (see \cite[Section 3.2.1]{CicDE:2017Submitted} for more details). Finally, the Dirichlet boundary conditions are enforced strongly, and the linear systems are solved using the direct solver PardisoLU from the MKL library (alternatively, iterative solvers are also applicable). Dunavant quadratures \cite{Dunavant1985} are used with order $2k$ for stabilized HHO methods, and with order $(2k+2)$ for unstabilized HHO methods.
\section{Test cases with known solution}\label{sec_numconv}
The goal of this section is to evaluate the stabilized and unstabilized
HHO methods on some test cases with known solution. This allows us to
compute errors on the displacement and the gradient as $\normev[\Omega_0]{\vu
  - \uTh}$ and $\normem[\Th]{\gradX \vu - \Grh(\uTh, \uFh)}$ where $\vu$
is the exact solution. 
We assess the convergence rates to smooth
solutions and we study the behavior of the HHO methods in the
quasi-incompressible regime. We consider two- and three-dimensional
settings. We use the abridged notation uHHO($k$) for the unstabilized
method with $\RecT = \Pkpd(T;\Reel^{d\times d})$ and sHHO($k$) with
$\RecT=\Pkd(T;\Reel^{d\times d})$ for the stabilized method; whenever
the context is unambiguous, we drop the polynomial degree $k$.
All the considered meshes are matching, simplicial affine meshes.
\subsection{Three-dimensional manufactured solution}\label{sec:manufactured}
We first report convergence rates for a nonlinear problem with a
manufactured solution in three space dimensions. We denote  by
$\Xo = (X,Y,Z)$ the Cartesian coordinates in $\Reel^3$. We set $\Gamma =
\Bd$ and the value of $\vu_d$ is determined from the exact solution on
$\Bd$. Concerning the constitutive relation, we take $\mu = 1$, $\lambda
=10$ (which corresponds to a Poisson ratio of $\nu \simeq 0.455$), and
$\Theta(J) = \ln J$. We consider the unit cube $\Omega_0 = (0,1) \times (0,1) \times (0,1)$ and the exact displacement solution is 
\begin{subequations}\begin{align}
u_X &= \left(\frac{1}{\lambda} + \alpha \right) X + \vartheta(Y), \quad u_Y = -\left(\frac{1}{\lambda} + \frac{\alpha + \gamma + \alpha\gamma}{1 + \alpha + \gamma + \alpha\gamma} \right) Y, \\
 \quad u_Z& = \left(\frac{1}{\lambda} + \gamma \right) Z + g(X) + h(Y), 
\end{align}\end{subequations}
where $\alpha$ and $\gamma$ are positive real numbers, and $\vartheta: \Reel \rightarrow \Reel$,  $g: \Reel \rightarrow \Reel$, $h: \Reel \rightarrow \Reel$ are smooth functions. Choosing $\vartheta(Y) =  \alpha  \sin(\pi Y)$, $g(X) =  \gamma  \sin(\pi X)$, and $h(Y) = 0$, the corresponding body forces are given by
 \begin{equation}
f_X = \mu \alpha \pi^2 \sin( \pi X), \qquad f_Y = 0, \qquad f_Z = \mu \gamma \pi^2 \sin( \pi Y).
\end{equation}
We set $\alpha = \gamma =0.1$. The stabilization parameter is taken as $\beta_0 = 1$ for sHHO. The displacement and gradient errors are reported as a function of the average mesh size $h$ for $k=1$ in Tab.~\ref{tab:rates_3d_k1}, for $k=2$ in Tab.~\ref{tab:rates_3d_k2} and for $k=3$ in Tab.~\ref{tab:rates_3d_k3}. For all $k \in \{1,2,3\}$, the displacement and the gradient converge, respectively, with order $(k+2)$ and $(k+1)$ for sHHO and with order $(k+1)$ and $k$ for uHHO. These convergence rates are consistent with the discussion at the end of Sections~\ref{sec:uHHO} and~\ref{sec:sHHO} on the convergence rates to be expected for linear elasticity and smooth solutions.
\begin{table}
\centering
\begin{tabular}{|c|c|c|c|c|c|c|c|c|}
\hline
Mesh & \multicolumn{4}{c|}{sHHO(1)} &\multicolumn{4}{c|}{uHHO(1)} \\ 
\cline{2-9} 
size & \multicolumn{2}{c|}{Displacement}  &\multicolumn{2}{c|}{Gradient} & \multicolumn{2}{c|}{Displacement} & \multicolumn{2}{c|}{Gradient} \\ 
\cline{2-9} 
$h$ & Error & Order & Error & Order & Error & Order & Error & Order \\ 
\hline 
 4.75e-1 &1.14e-3  & - &   9.40e-3 & - & 1.85e-3  & - & 6.64e-2  & - \\
\hline
 3.21e-1 & 4.27e-4 &  2.49& 4.66e-3& 1.78 & 7.76e-4 & 2.22 & 4.00e-2 &  1.25\\
\hline 
2.19e-1 & 1.22e-4 & 3.28  & 2.24e-3 & 1.91 &  3.49e-4 & 2.10 &2.95e-2  & 0.84 \\ 
\hline 
1.76e-1 & 6.36e-5 & 2.97 & 1.51e-3  & 1.79 & 2.19e-4  & 2.12 & 2.36e-2 & 1.01 \\ 
\hline 
1.39e-1 & 3.10e-5 & 3.05 & 9.16e-4 & 2.14 & 1.36e-4 & 2.01 & 1.88e-2  &  0.96\\ 
\hline 
1.11e-1 & 1.56e-5 & 3.00 & 5.92e-4 & 1.91 & 8.79e-5  &  1.94 & 1.50e-2  &  1.00\\ 
\hline 
\end{tabular}
\caption{3D manufactured solution: errors vs. $h$ for $k=1$.}
\label{tab:rates_3d_k1}
\bigskip
\begin{tabular}{|c|c|c|c|c|c|c|c|c|}
\hline
Mesh & \multicolumn{4}{c|}{sHHO(2)} &\multicolumn{4}{c|}{uHHO(2)} \\ 
\cline{2-9} 
size & \multicolumn{2}{c|}{Displacement}  &\multicolumn{2}{c|}{Gradient} & \multicolumn{2}{c|}{Displacement} & \multicolumn{2}{c|}{Gradient} \\ 
\cline{2-9} 
$h$ & Error & Order & Error & Order & Error & Order & Error & Order \\ 
\hline 
4.75e-1 & 1.04e-4  & -    & 9.89e-4 & - & 1.96e-4  & - & 7.68e-3 & - \\
\hline
3.21e-1 &  3.01e-5 & 3.16 & 3.18e-4 & 2.71  & 6.10e-5 & 2.96& 3.51e-3 &  1.96 \\
\hline 
2.19e-1 &  4.54e-6 & 4.04  & 9.57e-5 & 3.01 & 1.68e-5 & 3.17 & 1.60e-3 & 2.02  \\ 
\hline 
1.76e-1 &  1.79e-6 &  4.23 &  4.78e-5 & 3.16 & 9.72e-6 & 2.49 & 1.10e-3 & 1.68 \\ 
\hline 
1.39e-1 &  7.23e-7 &  3.85 & 2.35e-5 & 3.01 & 4.30e-6 & 3.36& 6.53e-4 & 2.24 \\ 
\hline 
1.11e-1 &  2.93e-7 &   3.96 & 1.21e-5  & 2.91 &  2.23e-6 & 2.88&  4.20e-4 & 1.94   \\ 
\hline 
\end{tabular}
\caption{3D manufactured solution: errors vs. $h$ for $k=2$.}
\label{tab:rates_3d_k2}
\bigskip
\begin{tabular}{|c|c|c|c|c|c|c|c|c|}
\hline
Mesh & \multicolumn{4}{c|}{sHHO(3)} &\multicolumn{4}{c|}{uHHO(3)} \\ 
\cline{2-9} 
size & \multicolumn{2}{c|}{Displacement}  &\multicolumn{2}{c|}{Gradient} & \multicolumn{2}{c|}{Displacement} & \multicolumn{2}{c|}{Gradient} \\ 
\cline{2-9} 
$h$ & Error & Order & Error & Order & Error & Order & Error & Order \\ 
\hline 
4.75e-1 & 7.39e-6 &  -     & 6.42e-5  & - &   1.59e-5 & -  & 7.79e-4 & - \\ 
\hline
3.21e-1 &  9.88e-7 &  5.13& 1.67e-5 &  3.41 & 2.80e-6 &  4.42 & 2.16e-4 & 3.26  \\
\hline 
2.19e-1 & 1.58e-7 &  4.79 & 2.98e-6 &  4.53  & 5.23e-7 &  4.40 & 6.55e-5 & 3.14  \\ 
\hline 
1.76e-1 &  5.54e-8 &  4.77 & 1.37e-6 & 3.52 & 2.07e-7 &  4.21 & 3.23e-5 & 3.19   \\ 
\hline 
1.39e-1 & 1.60e-8  &  5.29 & 4.86e-7 &  4.43 & 8.08e-8 &  4.01 & 1.61e-5 &  2.95   \\ 
\hline
1.11e-1 & 5.01e-9  &  5.17  & 1.96E-7 & 4.03 & 3.25e-8  & 4.05 &  8.25E-6 & 2.97  \\ 
\hline 
\end{tabular}
\caption{3D manufactured solution: errors vs. $h$ for $k=3$.}
\label{tab:rates_3d_k3}
\end{table}
\subsection{Quasi-incompressible annulus} \label{sec:annulus}
Our goal is now to evaluate the sHHO and uHHO methods in the
quasi-incompressible case for finite deformations. We consider a test
case from \cite[Section~5.2]{KaLeC:2015}
that consists of an annulus centered at the origin
with inner radius $R_0 = 0.5$ and outer
radius $R_1=1$. The annulus is deformed by imposing a displacement $\vu_d(\Xo)
= \Xo (r_0 -R_0)/R_0$ on $\Bd = S_{R_0}$ where $r_0$ is a real positive
parameter, and $\Tn = \vecteur{0}$ on $\Bn =S_{R_1}$ ($S_R$ is the
sphere of radius $R$ centered at the origin). An accurate reference
solution can be computed by solving an ordinary 
differential equation along the radial 
coordinate, as detailed in \cite{KaLeC:2015}.
We set $r_0 = 1.5$ and $\mu = 0.333$ (different values of $\lambda$ are
considered). Since we use meshes with planar faces, we only consider $k=1$. 

\begin{figure}[htbp]
    \centering
    \subfloat[Reference and deformed configuration]{
    \label{fig:def_annulus_a}
        \centering 
        \includegraphics[scale=0.27]{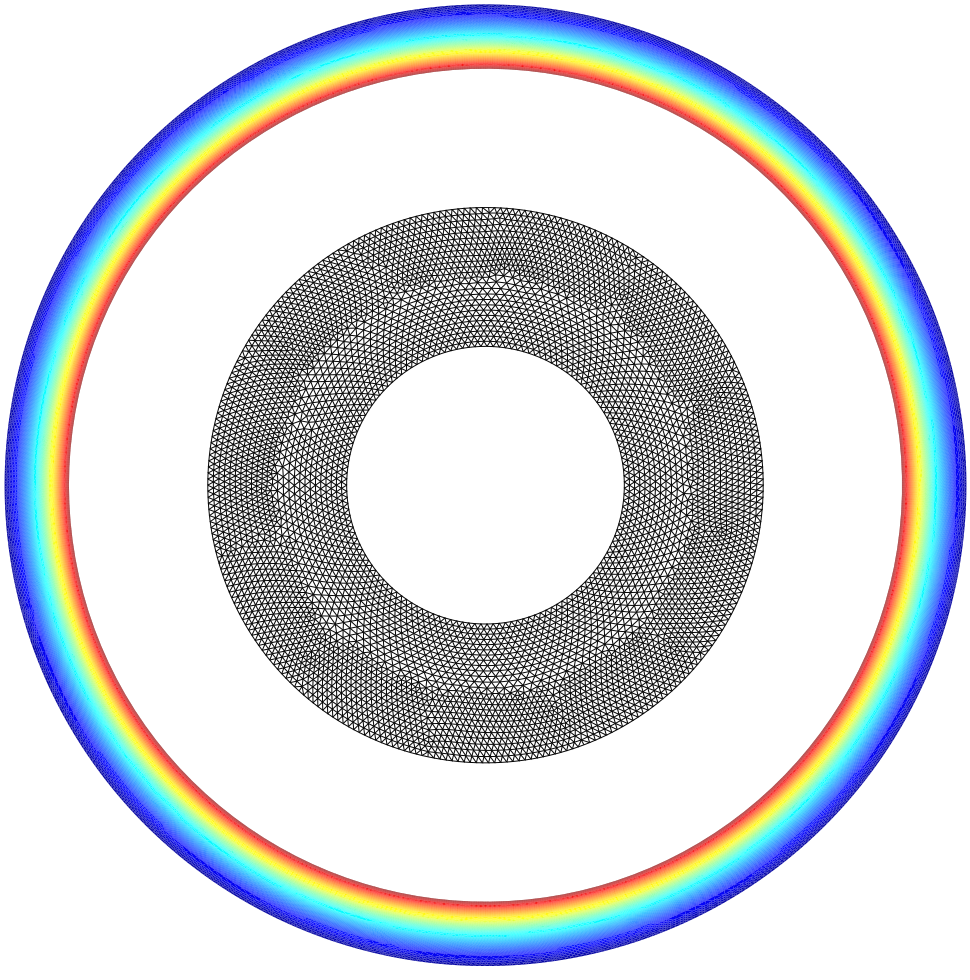} 
  }
    ~ 
    \subfloat[Discrete Jacobian on the reference configuration]{
    \label{fig:def_annulus_b}
        \centering
         \raisebox{10mm}{\includegraphics[scale=0.13]{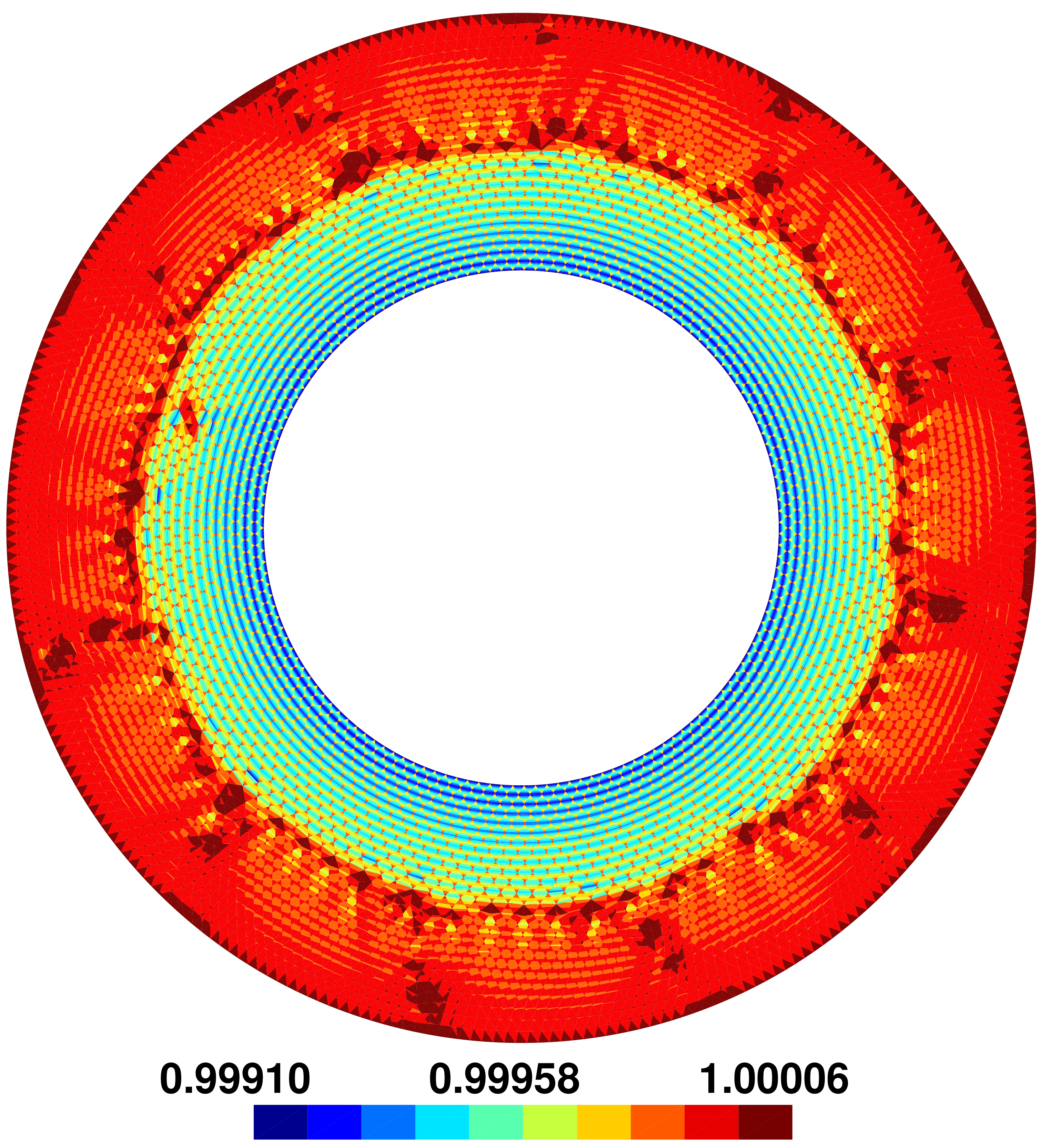} }
        
    }
    \caption{Quasi-incompressible annulus with $\lambda=1666.44$: sHHO(1) solution on a mesh composed of 10161 triangles.}\label{fig:def_annulus}
\end{figure}

\begin{figure}[htbp]
    \centering
    \subfloat[Displacement error]{
        \centering 
\includegraphics[scale=1.0]{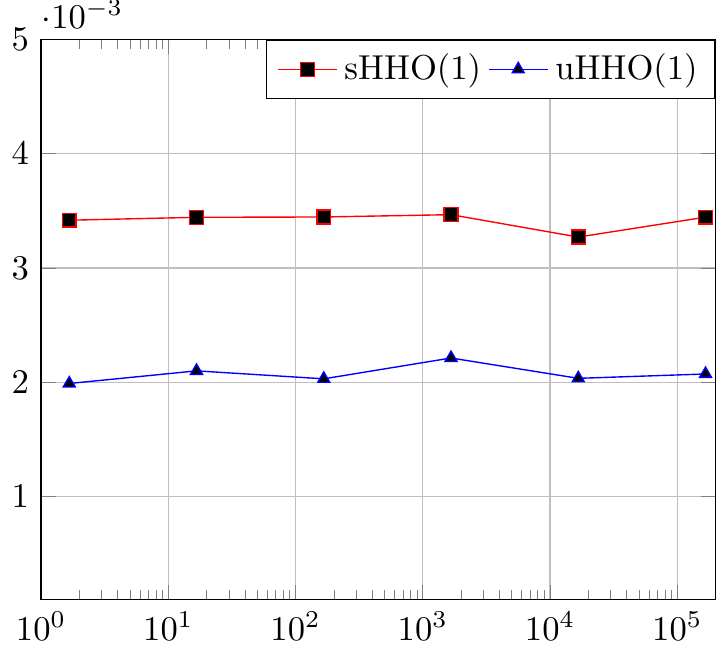} 
    }
    ~ 
    \subfloat[Gradient error]{
        \centering
\includegraphics[scale=1.0]{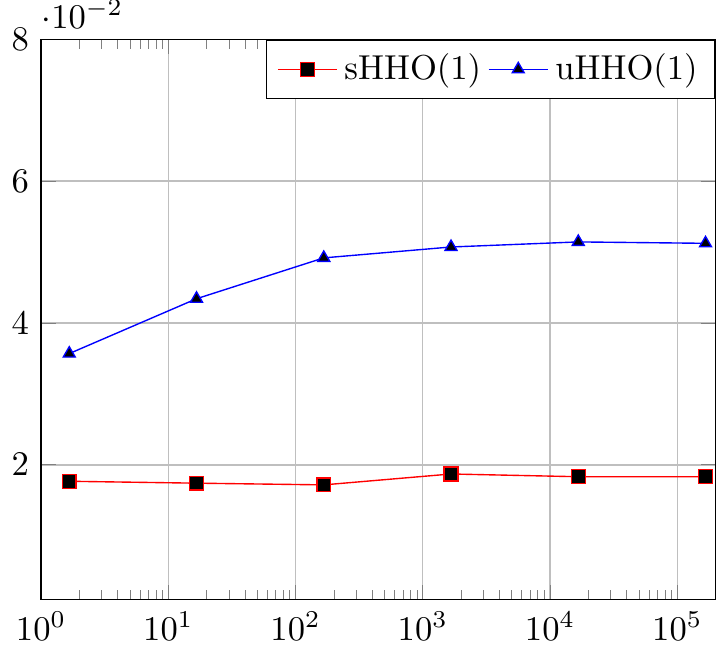} 
    }
    \caption{Quasi-incompressible annulus: errors vs. $\lambda$ for $h=\text{2.52e-2}$}\label{fig:locking_annulus}
\end{figure}

\begin{figure}[htbp]
    \centering
    \subfloat[sHHO(1): $u_r$ vs. $r$]{
        \centering 
        \includegraphics[scale=0.95]{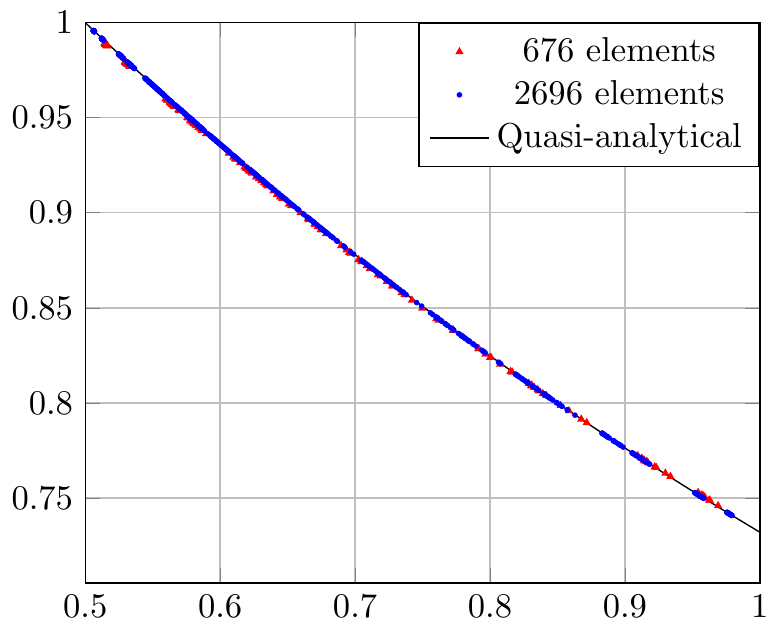} 
    }
    ~ 
    \subfloat[uHHO(1): $u_r$ vs. $r$]{
        \centering
        \includegraphics[scale=0.95]{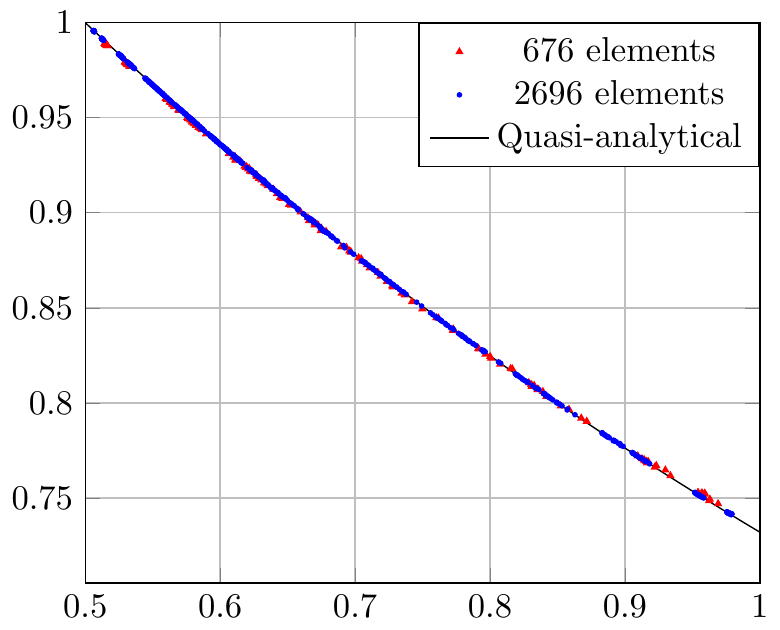} 
    }
    
       \subfloat[sHHO(1): $P_{rr}$ vs. $r$]{
        \centering 
        \includegraphics[scale=0.945]{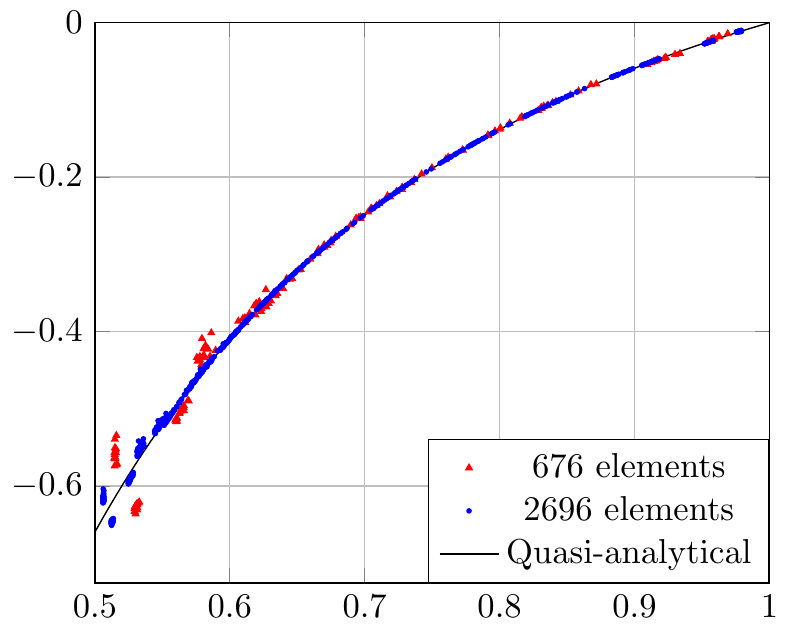} 
    }
    ~ 
    \subfloat[uHHO(1): $P_{rr}$ vs. $r$]{
        \centering
        \includegraphics[scale=0.945]{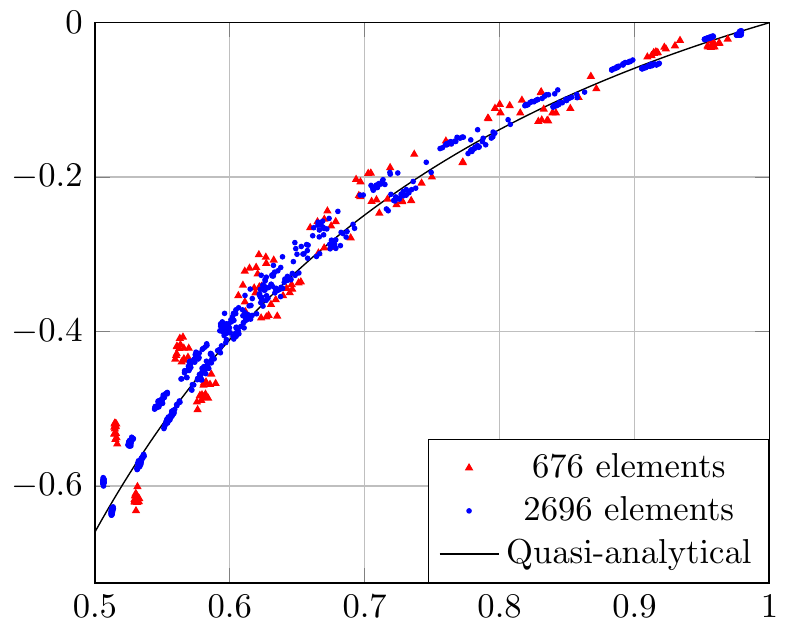} 
    }
    
        \subfloat[sHHO(1): $P_{\theta\theta}$ vs. $r$]{
        \centering 
        \includegraphics[scale=0.95]{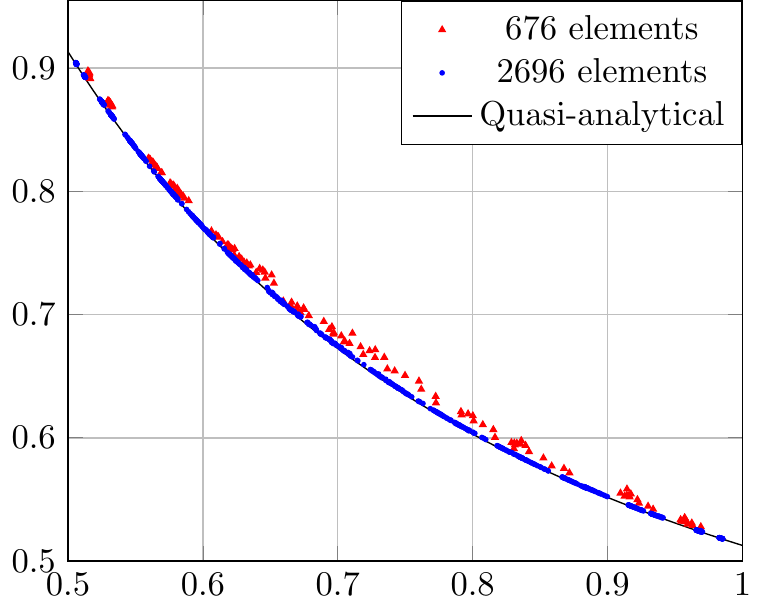} 
    }
    ~ 
    \subfloat[uHHO(1): $P_{\theta\theta}$ vs. $r$]{
        \centering
        \includegraphics[scale=0.95]{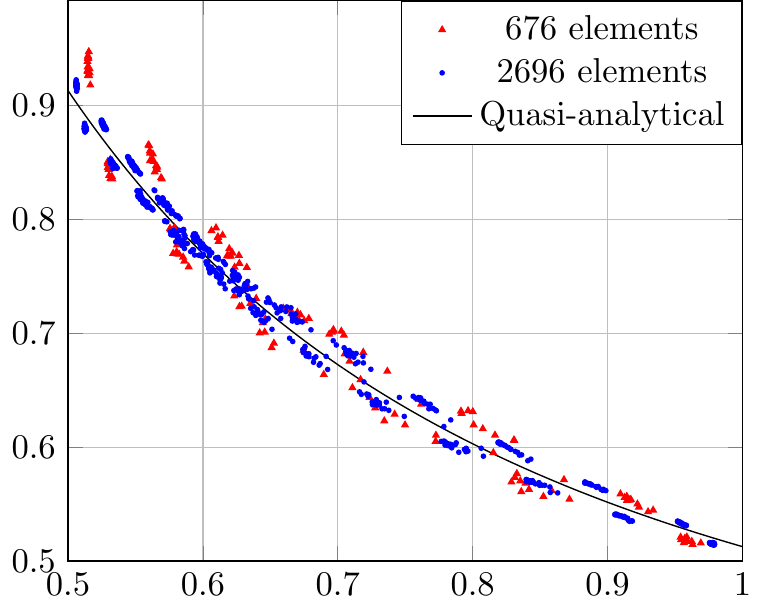} 
    }
\caption{Quasi-incompressible annulus with $\lambda=1666.44$: comparison of the reference and computed values of $u_r$, $P_{rr}$ and $P_{\theta\theta}$ at the barycenter of the mesh cells (located in the upper quadrant) for two different meshes obtained with the sHHO and uHHO methods.}
\label{fig::sol_pt_annulus}
\end{figure}

The reference and deformed configuration for sHHO(1) are
shown in Fig.~\ref{fig:def_annulus_a} for $\lambda =
1666.44$ (which corresponds to a Poisson ratio of $\nu \simeq 0.4999$).
The stabilization parameter has to be of the order of $\beta_0=100$ 
to achieve convergence.
In Fig.~\ref{fig:def_annulus_b}, 
we display the discrete Jacobian $J^h$ on the reference configuration (computed using sHHO(1)), and we observe
that this quantity takes values very close to 1 everywhere in the
annulus (as expected). Convergence rates for the displacement and the
gradient are reported in Tab.~\ref{tab:rates_annulus_k1} for $\lambda =
1666.44$ (similar convergence rates, not reported herein, are observed
for lower values of $\lambda$). We observe that for sHHO, the displacement
and the gradient converge with order 2, whereas for uHHO, the
displacement converges with order 2 and the gradient with order
1. More importantly, the errors are uniform with respect to $\lambda$ as
shown Fig.~\ref{fig:locking_annulus}. This result confirms numerically
that in this case, sHHO and
uHHO remain locking-free in quasi-incompressible finite
deformations. Incidentally, we notice that sHHO produces slightly lower
errors than uHHO which is consistent with the higher-order convergence for sHHO.
Moreover, the displacement on the boundary is imposed by uniform load
increments. For $\lambda = 1666.44$, sHHO requires 30 loading steps with
a total of 125 Newton's iterations, whereas uHHO requires 33 loading
steps with a total of 137 Newton's iterations, i.e., sHHO is about 10\%
more computationally-effective than uHHO in this example. Finally, the reference values of $u_r$, $P_{rr}$ and $P_{\theta\theta}$ at the barycenter of each cell are plotted in Fig.~\ref{fig::sol_pt_annulus} for $\lambda=1666.44$, showing the pointwise convergence of the various discrete solutions. 
We observe that for both HHO methods, the error on $P_{rr}$ is slightly more important near the inner boundary of the annulus (where the stress is maximal).
\begin{table}
\centering
\begin{tabular}{|c|c|c|c|c|c|c|c|c|}
\hline
Mesh & \multicolumn{4}{c|}{sHHO(1)} &\multicolumn{4}{c|}{uHHO(1)} \\ 
\cline{2-9} 
size & \multicolumn{2}{c|}{Displacement}  &\multicolumn{2}{c|}{Gradient} & \multicolumn{2}{c|}{Displacement} & \multicolumn{2}{c|}{Gradient} \\ 
\cline{2-9} 
$h$ & Error & Order & Error & Order & Error & Order & Error & Order \\ 
\hline 
1.15e-1 & 5.98e-2   & - & 3.22e-1 & - & 4.00e-2 & - & 1.23e-1  & - \\ 
\hline 
5.77e-2& 1.81e-2 & 1.72 & 8.23e-1 & 1.97 & 1.32e-2 & 1.62  & 1.01e-1 & 0.28 \\ 
\hline 
3.45e-2 & 6.30e-3 & 2.05 & 3.15e-2 & 1.86 & 3.80e-3  & 2.42 & 6.60e-2 & 0.83 \\ 
\hline 
2.52e-2 & 3.42e-3 & 1.95 & 1.83e-2 & 1.73 & 2.03e-3 & 2.05 & 5.11e-2 & 0.94 \\ 
\hline 
1.64e-2 & 1.49e-3 & 1.93 & 7.98e-3 & 1.93 & 9.76e-4 & 1.72 & 3.09e-2 & 1.08 \\ 
\hline 
\end{tabular}
\caption{Quasi-incompressible annulus: errors vs. $h$ for $k=1$ and $\lambda = 1666.44$. }
\label{tab:rates_annulus_k1}
\end{table}
\subsection{Efficiency}
In this section, we compare the performance of sHHO, uHHO and that of a
continuous Galerkin (cG) method in terms of efficiency when solving the
three-dimensional manufactured solution from
Section~\ref{sec:manufactured}. The number of unknowns is the number of
degrees of freedom attached to faces after static condensation for sHHO
and uHHO and the number of degrees of freedom attached to nodes for
cG. The cG method is based on a primal formulation realized within the industrial open-source FEM software \CA~\cite{CodeAster} interfaced with the open-source \texttt{mfront} code generator~\cite{Mfront} to generate Neohookean laws.

\begin{figure}[htbp]
    \centering
    \subfloat[Displacement error vs. number of degrees of freedom]{
    \label{fig:tmp_dof}
        \centering 
\includegraphics[scale=0.9]{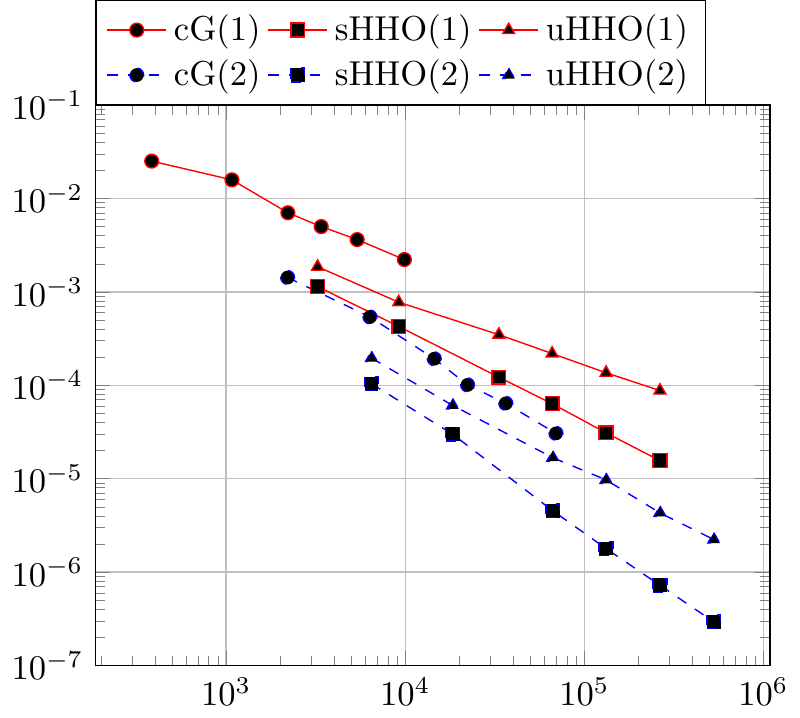} 
    }
    ~ 
    \subfloat[Displacement error vs number of non-zero entries in the stiffness matrix]{
    \label{fig:tmp_entries}
        \centering
\includegraphics[scale=0.9]{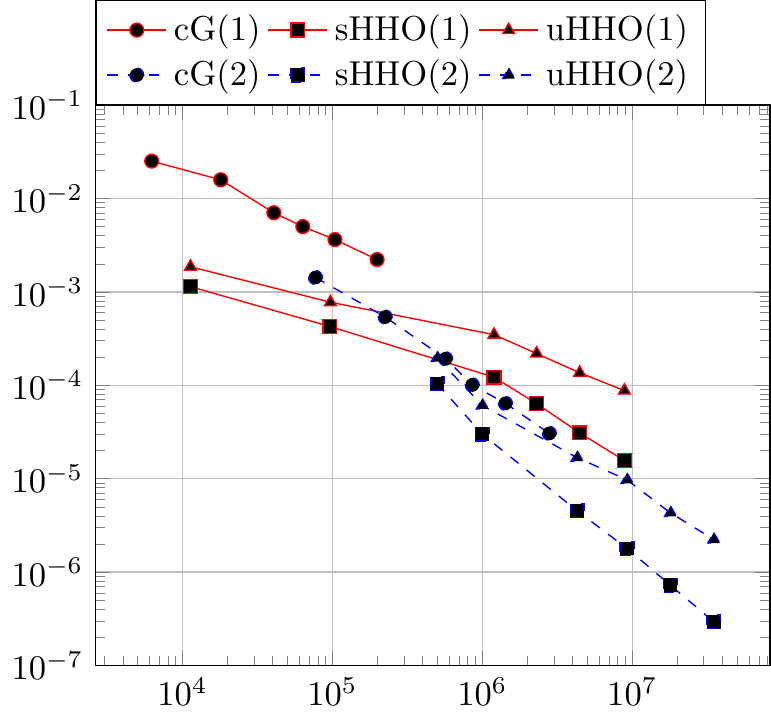} 
    }
\caption{3D manufactured solution: comparison of the displacement error
  obtained with sHHO, uHHO, and cG.}
\label{tmp-sum}
\end{figure}

\begin{figure}[htbp]
    \centering
    \subfloat[$\tau_{\rm ass}$ / $\tau_{\rm sol}$ vs. card($\Fh$)]{
        \centering 
\includegraphics[scale=0.9]{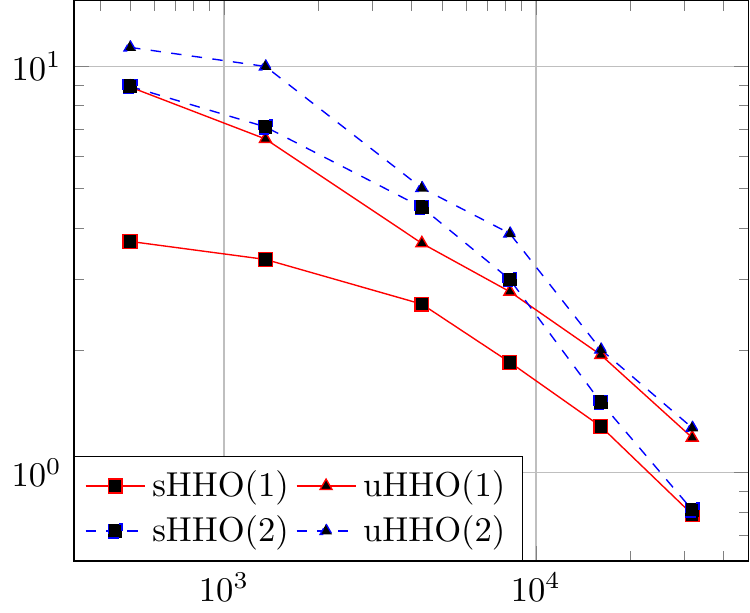} 
		\label{fig:tmp_systeme}
  }
    ~ 
    \subfloat[Time for the different operations normalized by the total time for sHHO(1) for a mesh with 31621 faces]{
        \centering
\includegraphics[scale=0.9]{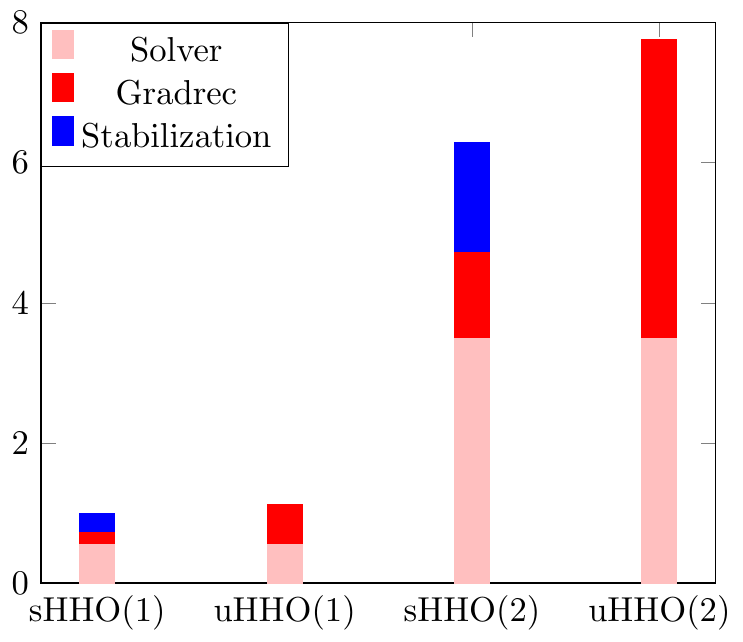} 
        \label{fig:tmp_op}
    }
    \caption{Comparison of CPU times for the sHHO and uHHO methods.}
\end{figure}

We present the displacement error versus the number of degrees of
freedom in Fig.~\ref{fig:tmp_dof} and versus the number of non-zero
entries in the stiffness matrix in Fig.~\ref{fig:tmp_entries}. Owing to
the static condensation, we observe that, for the same approximation order
and the same number of degrees of freedom or non-zero entries 
in the stiffness matrix, the
displacement error is smaller for sHHO than for cG and comparable between uHHO and cG.  
Let us now compare the time spent to solve the non-linear problem when using
sHHO($k$) and uHHO($k$) with $k\in\{1,2\}$. 
For the present test case, the nonlinear problem is
solved, for both methods, in four Newton's iterations.
The codes are instrumented to measure the assembly time $\tau_{ass}$ to
build the local contributions to the global stiffness matrix and the
solver time  $\tau_{sol}$ which corresponds to solving the global linear
system ($\tau_{ass}$ and $\tau_{sol}$ are computed after summation
over all the Newton's steps). In \texttt{DiSk++}, the linear algebra
operations are realized using the Eigen library and the global
linear system (involving face unknowns only)  is solved with
PardisoLU. The tests are run sequentially on a 3.4 Ghz Intel Xeon
processor with 16 Gb of RAM.  In Fig.~\ref{fig:tmp_systeme} we plot the
ratio $\tau_{\rm ass} /  \tau_{\rm sol}$ versus the number of mesh faces, 
card$(\Fh)$. We can see that on
the finer meshes, the cost of local computations becomes negligible
compared to that of the linear solver; we notice that the situation is a
bit less favorable than for the results on linear elasticity reported in
\cite{DiPEr:2015} since the space to reconstruct the gradient is now
larger. In Fig.~\ref{fig:tmp_op} we provide a more detailed assessment of the cost on a fixed mesh with 31621 faces. More precisely, the time $\tau_{\rm ass}$ spent in assembling the problem is now divided into two parts, one part, denoted \texttt{Gradrec}, to reconstruct the gradient and build the global system to solve (the part related to static condensation is not included and takes a marginal fraction of the cost), and another part, denoted \texttt{Stabilization}, to build the stabilization operator for the sHHO method (including the time to build the displacement reconstruction, see \eqref{eq_reconstruction_depl}). In addition, the time $\tau_{\rm sol}$ spent in solving the system is now denoted \texttt{Solver}. We observe that the difference between sHHO($k$) and uHHO($k$) is not really important; in fact, the time that uHHO($k$) spends in reconstructing the gradient in a larger space is more or less equivalent to the time that sHHO($k$) spends in building the stabilization operator. 
Moreover, if memory is not a limiting factor, the gradient and the stabilization can be
computed once and for all, and re-used at each Newton's step. 

\begin{figure}[htbp]
\centering
\subfloat[Ratio of the condition number without and with static condensation vs. card($\Fh$)]{
\label{fig::cond_comp1}
\centering
\includegraphics[scale=0.9]{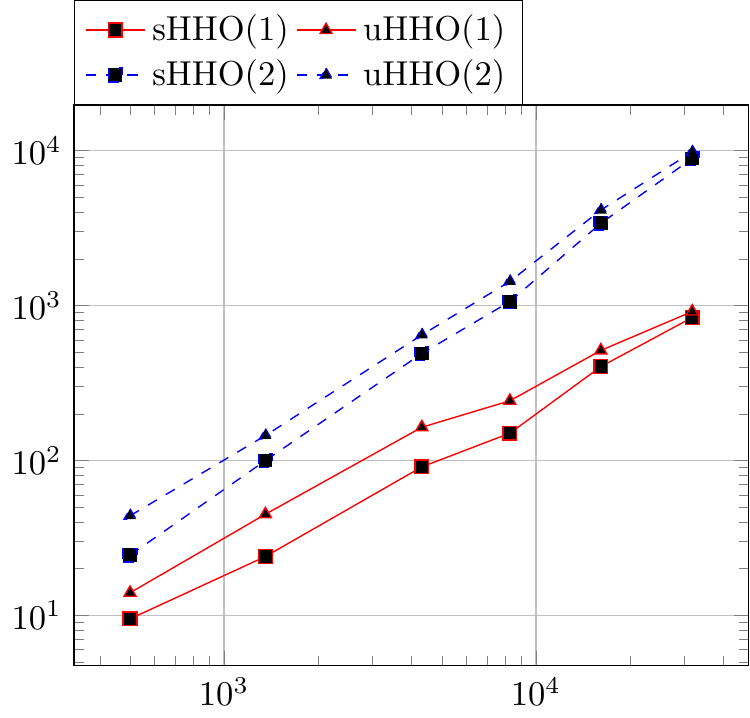} 
}
~ 
\subfloat[Condition number normalized by that for sHHO(1) or sHHO(2) with $\beta=1$ vs. card($\Fh$)]{
\label{fig::cond_comp2}
\centering
\includegraphics[scale=0.9]{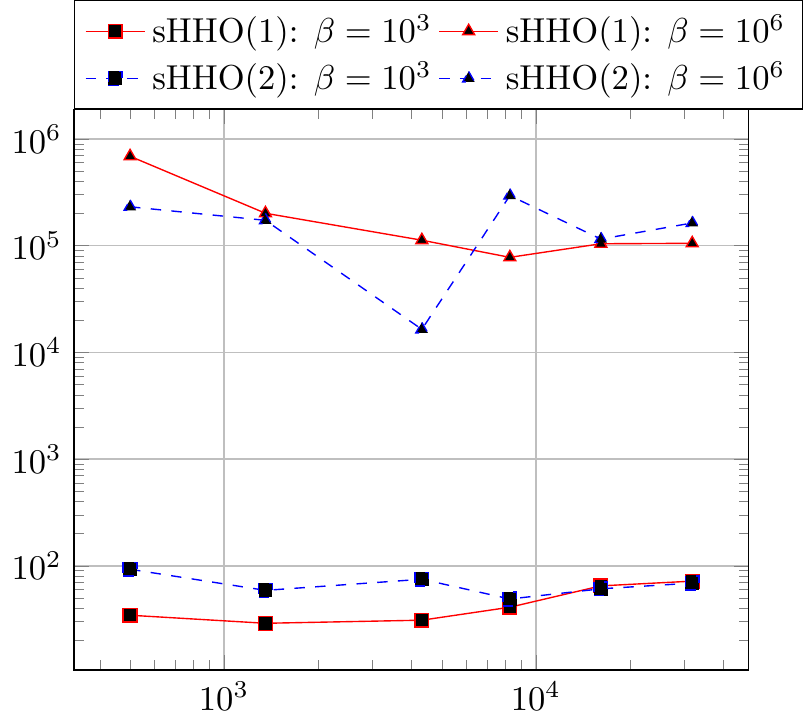} 
}
\caption{Condition number of the global stiffness matrix.}
\label{fig::cond_comp}
\end{figure}

Another interesting observation is that the condition number of the global stiffness matrix for both methods is improved by static condensation, as shown in Fig.~\ref{fig::cond_comp1} where the ratio of the condition number without and with static condensation is displayed as a function of the number of face degrees of freedom. This positive effect is even increased as the mesh is refined, and it is also more pronounced when the polynomial degree $k$ is higher. Finally, we assess the influence of the stabilization parameter $\beta$ on the condition number of the stiffness matrix for sHHO($k$) $k \in \{ 1,2\}$. Fig.~\ref{fig::cond_comp2} reports the condition number for $\beta\in \{10^3,10^6\}$ normalized by the condition number for $\beta=1$, as a function of the total number of face degrees of freedom. We observe that the condition number is amplified by a factor of $10^2$ when $\beta$ goes from 1 to $10^3$ and by a factor $10^3$ when $\beta$ goes from $10^3$ to $10^6$, independently of the polynomial degree $k$.

\section{Application-driven three-dimensional examples} \label{sec:3D}
The goal of this section is to show that sHHO and uHHO are capable of
dealing with challenging three-dimensional
examples with finite deformations. For the first test case, we compare
our results to those obtained 
with a cG method implemented in the industrial software \CA. For the
second and third test cases, we compare our results 
with the HDG solutions reported in
\cite{KaLeC:2015}. In all cases, we choose $\Theta(J) = \ln J$.
\subsection{Quasi-incompressible indented block}
\begin{figure}[htbp]
    \centering
    \subfloat[Euclidean displacement norm for cG(1) in the compressible regime]{
        \label{fig:bloc_def_cg1_comp}
        \centering
        \includegraphics[scale=0.25]{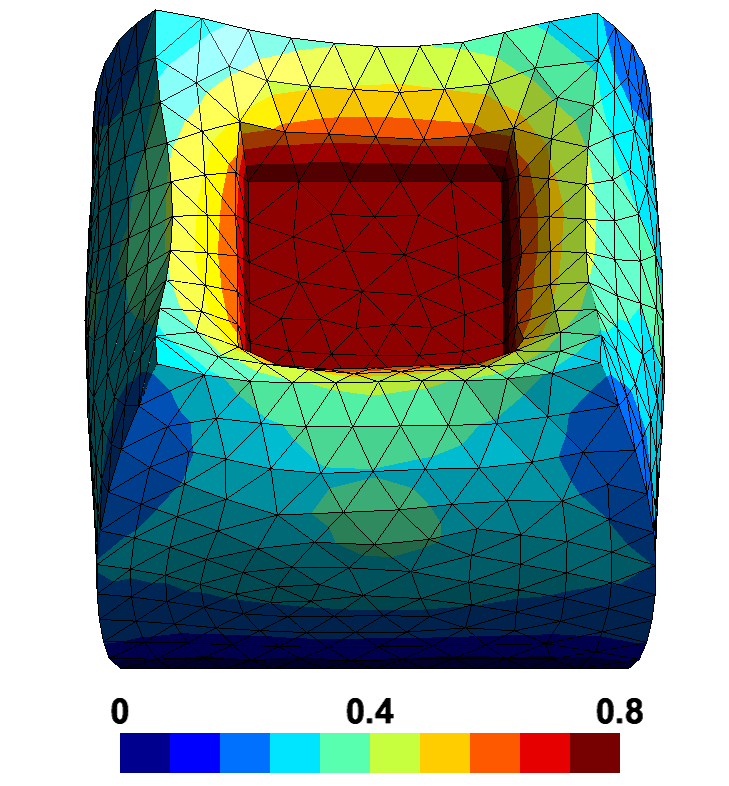} 
  }
    ~ 
    \subfloat[Euclidean displacement norm for sHHO(1) in the compressible regime]{
     \label{fig:bloc_def_shho1_comp}
        \centering
        \includegraphics[scale=0.25]{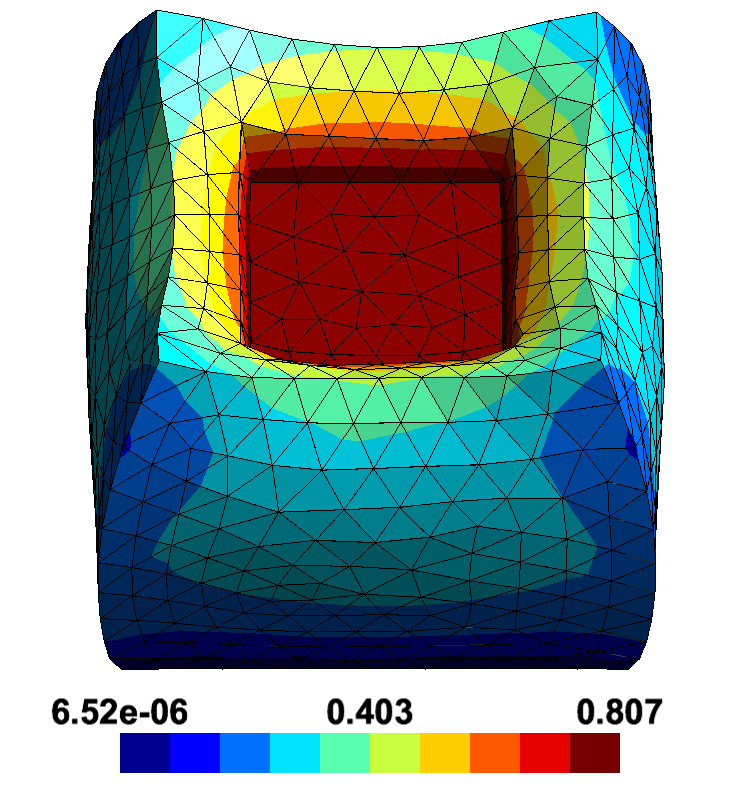}
    }
    
     \subfloat[Euclidean displacement norm for cG(1) in the quasi-incompressible regime]{
         \label{fig:bloc_def_cg1_incomp}
        \centering
        \includegraphics[scale=0.25]{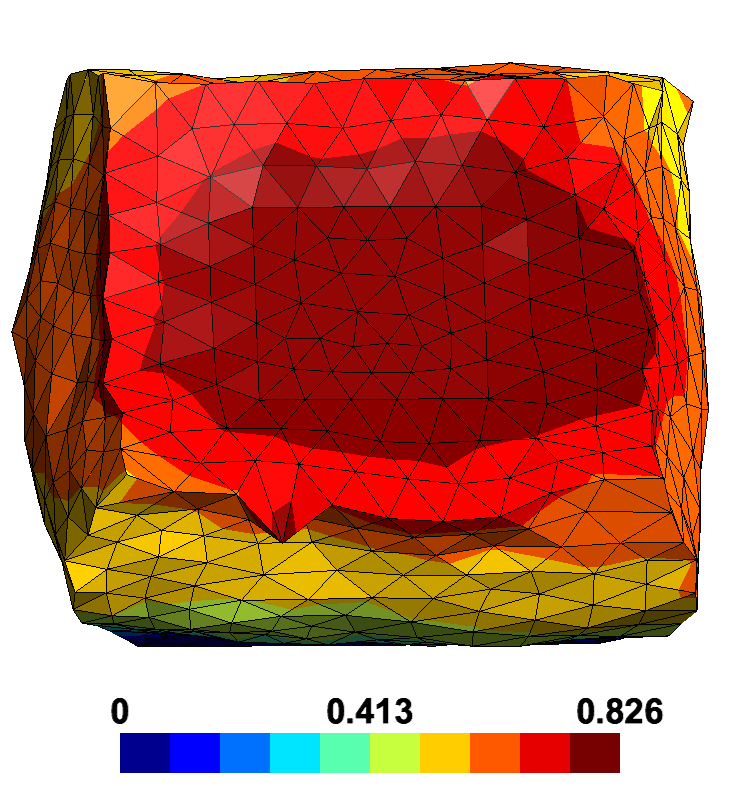}
  }
    ~ 
    \subfloat[Euclidean displacement norm for sHHO(1) in the quasi-incompressible regime]{
    \label{fig:bloc_def_shho1_incomp}
        \centering
        \includegraphics[scale=0.25]{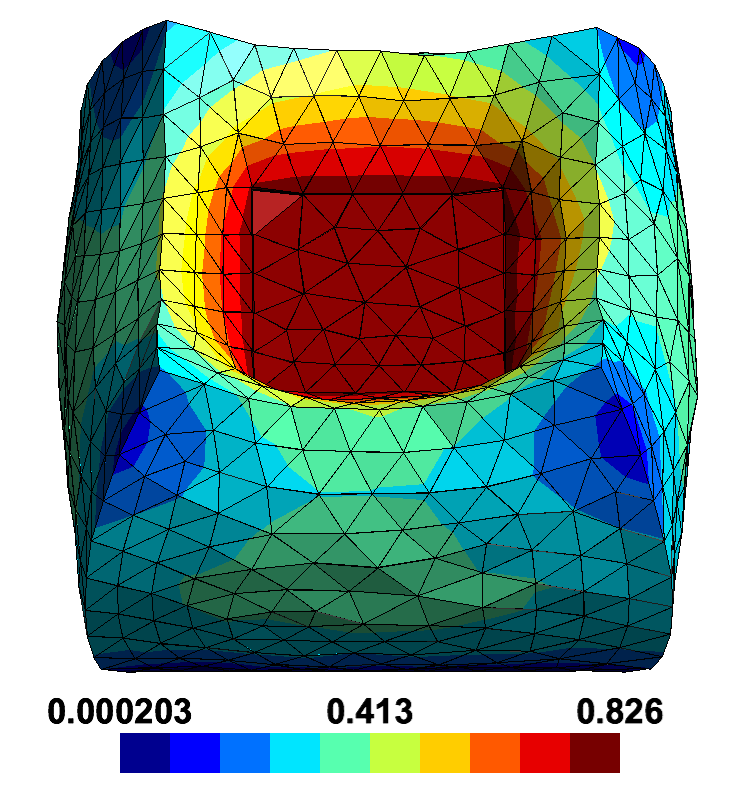}
    }
    \caption{Indented block: compressible (top) and
      quasi-incompressible regime (bottom) with Euclidean displacement norm shown in color on a mesh composed of 5526 tetrahedra.}\label{fig:bloc_def}
\end{figure}

\begin{figure}[htbp]
    \centering
    \subfloat[Compressible block]{
        \centering 
\includegraphics[scale=0.9]{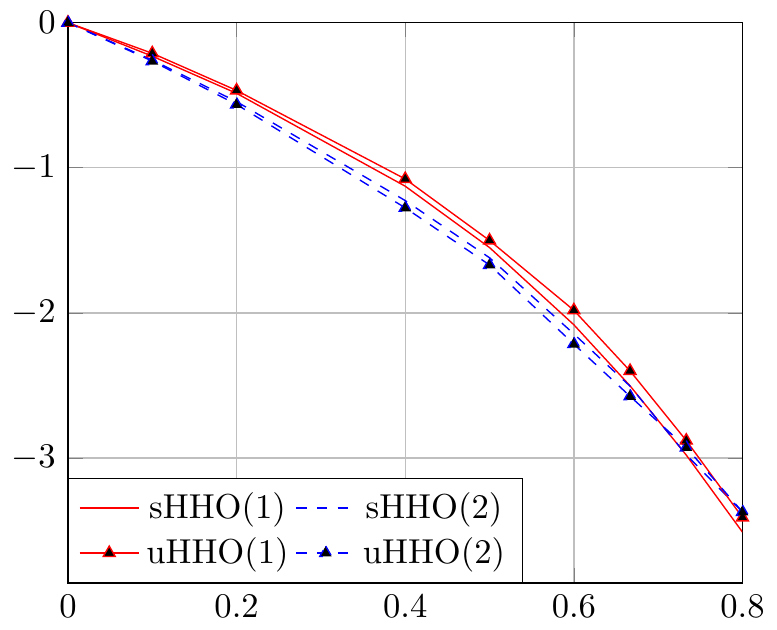}
    }
    ~ 
    \subfloat[Quasi-incompressible block]{
        \centering
\includegraphics[scale=0.9]{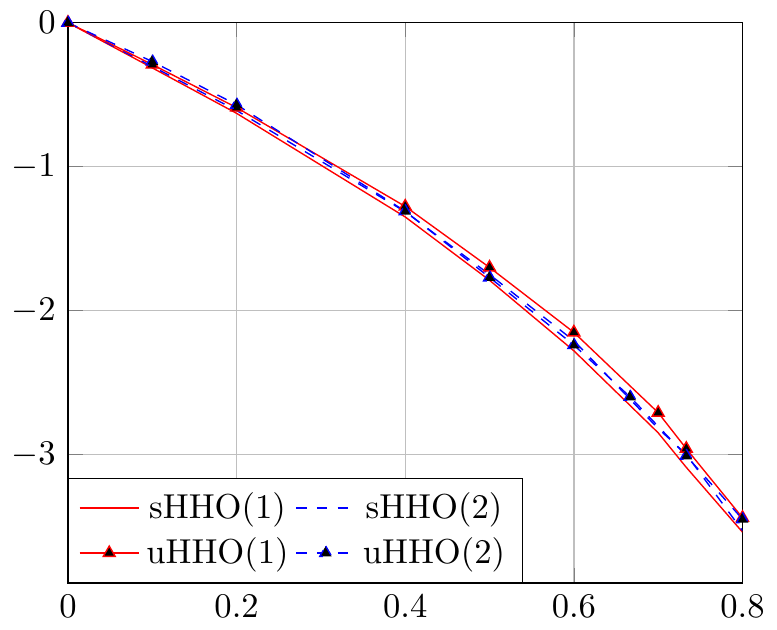}
    }
\caption{Indented block: vertical component of the computed discrete traction integrated over the indented surface versus the imposed vertical displacement using the sHHO and uHHO methods.}
  \label{fig::load_bloc}
\end{figure}

In this example, we model an indentation problem as a prototype for a
contact problem. We consider the unit cube $(-1,1) \times (-1,1) \times
(-1,1)$. To model the rigid indentor, the bottom surface is clamped, a vertical
displacement of $-0.8$ is imposed on the subset $(-0.5,0.5) \times (-0.5,0.5) \times\{ 1 \}$ of the top surface, and the other parts of the boundary are
traction-free. We set $\mu=1$ and $\lambda=4999$ in the quasi-incompressible regime (which corresponds to a Poisson ratio of $\nu \simeq 0.4999$). The stabilization parameter needs to be taken of the order of $\beta_0 = 100$ for sHHO. Fig.~\ref{fig:bloc_def_cg1_incomp} and Fig.~\ref{fig:bloc_def_shho1_incomp} present the Euclidean displacement norm on the deformed configuration obtained with cG(1) and sHHO(1) respectively (the uHHO(1) solution is very close to the sHHO(1) solution). 
We observe the locking phenomenon affecting the cG solution. To better
appreciate the influence of the parameter $\lambda$ on the discrete solutions,
we plot in Fig.~\ref{fig:bloc_def_cg1_comp} and Fig.~\ref{fig:bloc_def_shho1_comp} the Euclidean displacement norm on the deformed configuration in the compressible regime ($\lambda=1$, which corresponds to a Poisson ratio of $\nu = 0.25$).
We observe that in the compressible regime, the results produced by the various numerical methods are all very close, whereas the cG solutions depart from the the sHHO and uHHO solutions in the quasi-incompressible regime. Finally, the computed vertical component of the discrete traction integrated over the indented top surface is plotted in Fig.~\ref{fig::load_bloc} for sHHO and uHHO as a function of the imposed vertical displacement. The two HHO methods produce very similar results and capture well the nonlinear response of the block.
\subsection{Cylinder under compression and shear}
\begin{figure}[htbp]
    \centering
    \subfloat{
        \centering 
         \includegraphics[scale=0.2]{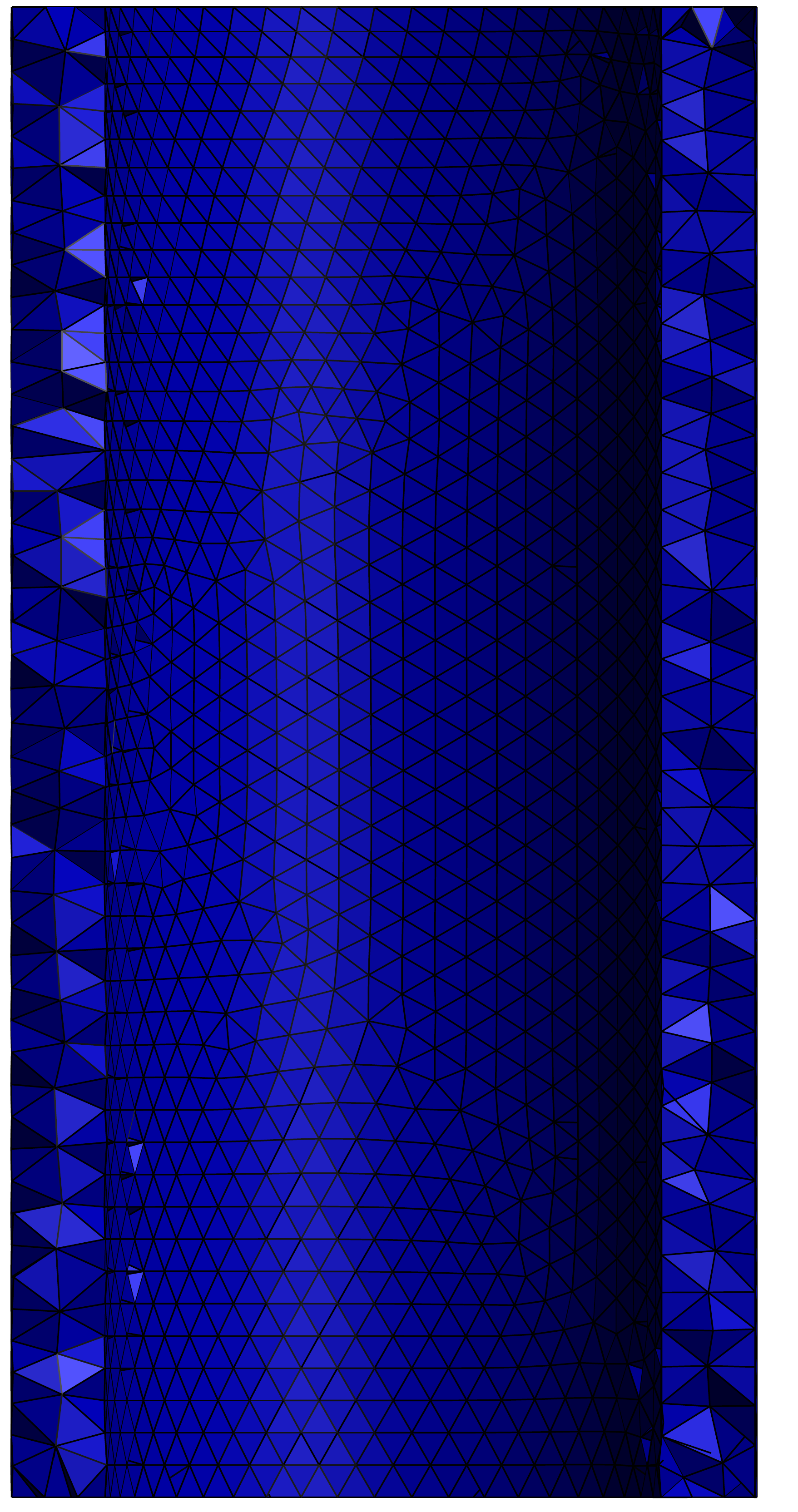}
    }
    ~ 
    \subfloat{
        \centering
         \includegraphics[scale=0.2]{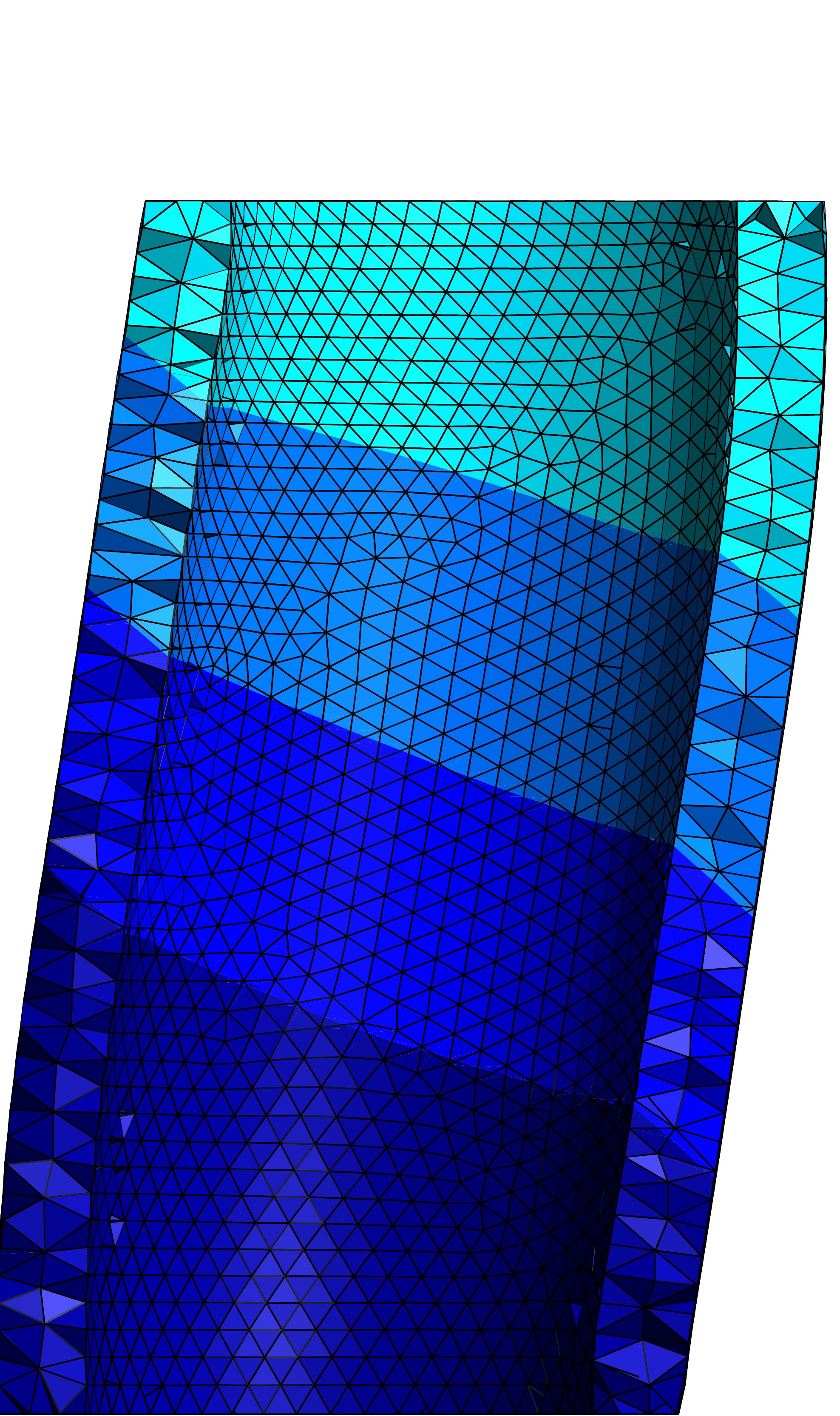}
    }
    ~ 
    \subfloat{
        \centering
        \includegraphics[scale=0.2]{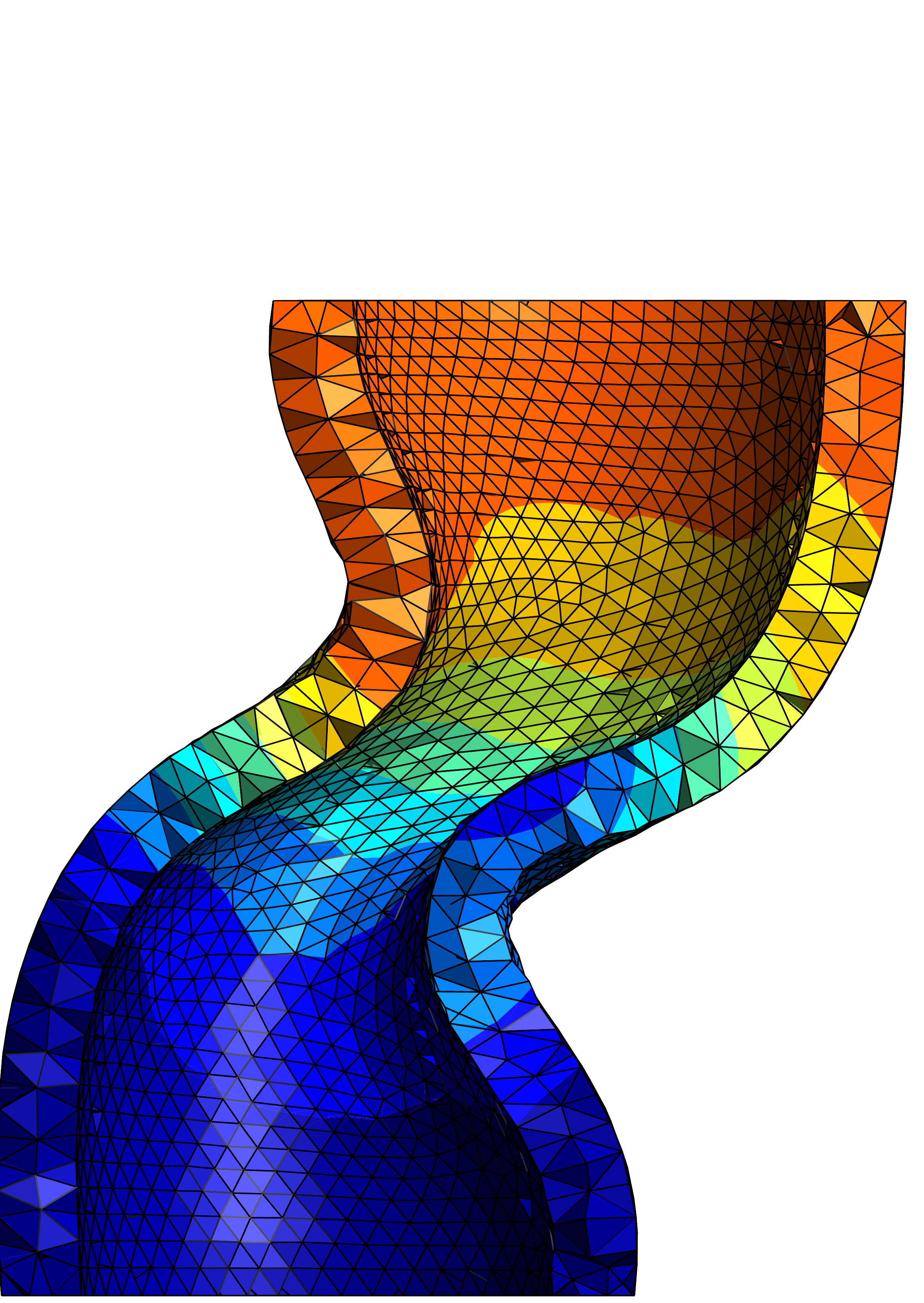}
    }

    \subfloat{
        \centering
        \includegraphics[scale=0.2]{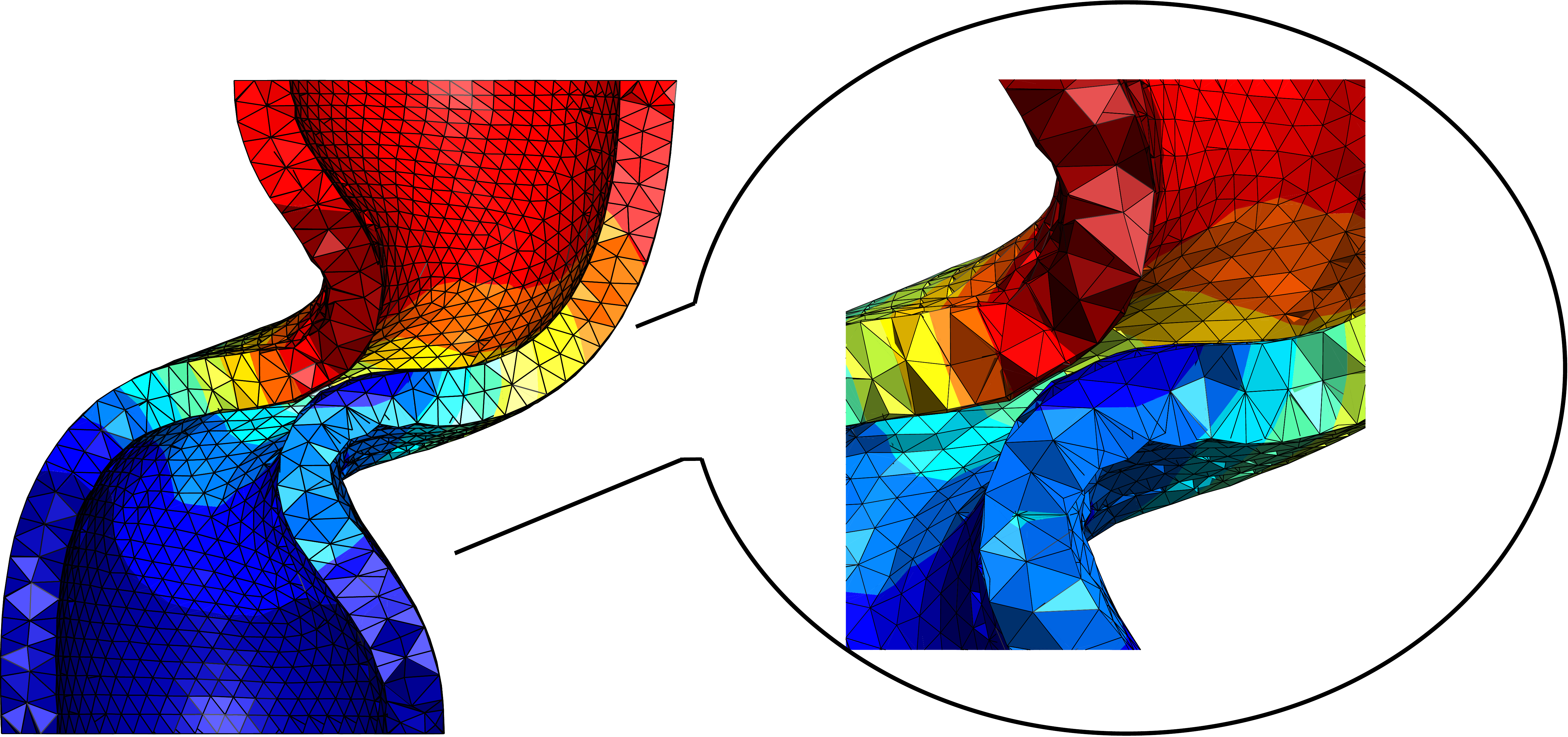}
    }
    \caption{Sheared cylinder: snapshots of the Euclidean displacement
      norm on the deformed configuration at $0\%$, $40\%$, $80\%$, and
      $100\%$ of loading, and a zoom where the deformations are the most
      important (uHHO(1) solution). The color scale goes from $0.0$ (blue) to $1.8$ (red).}\label{fig:def_cylindre}
\end{figure}

\begin{figure}[htbp]
    \centering
    \subfloat{
        \centering 
        \includegraphics[scale=0.25]{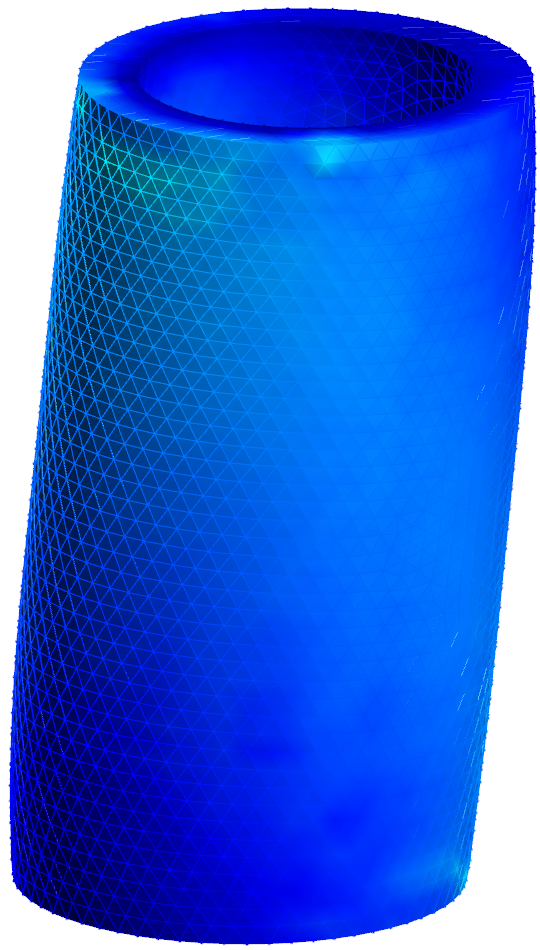} 
   }
    ~ 
    \subfloat{
        \centering
        \includegraphics[scale=0.25]{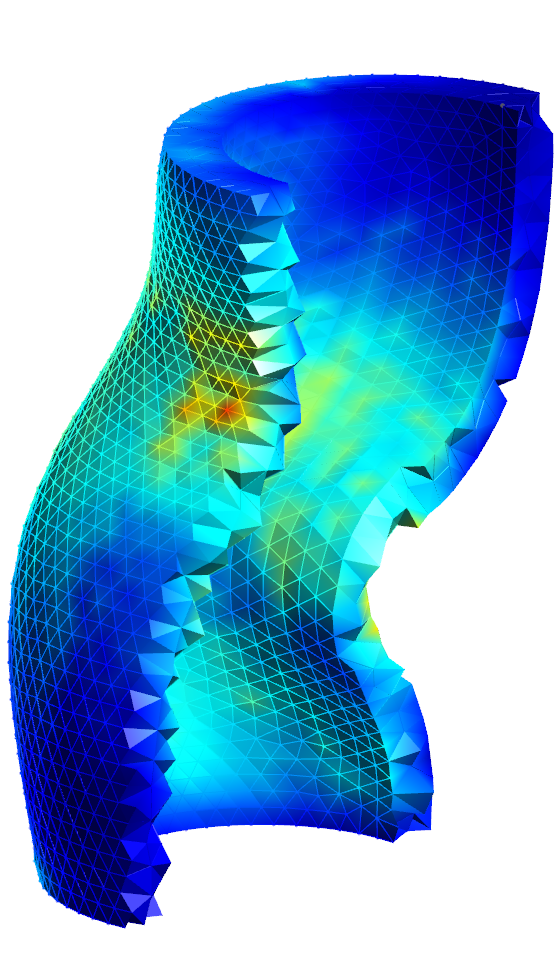} 
    }
    ~ 
    \subfloat{
        \centering
        \includegraphics[scale=0.25]{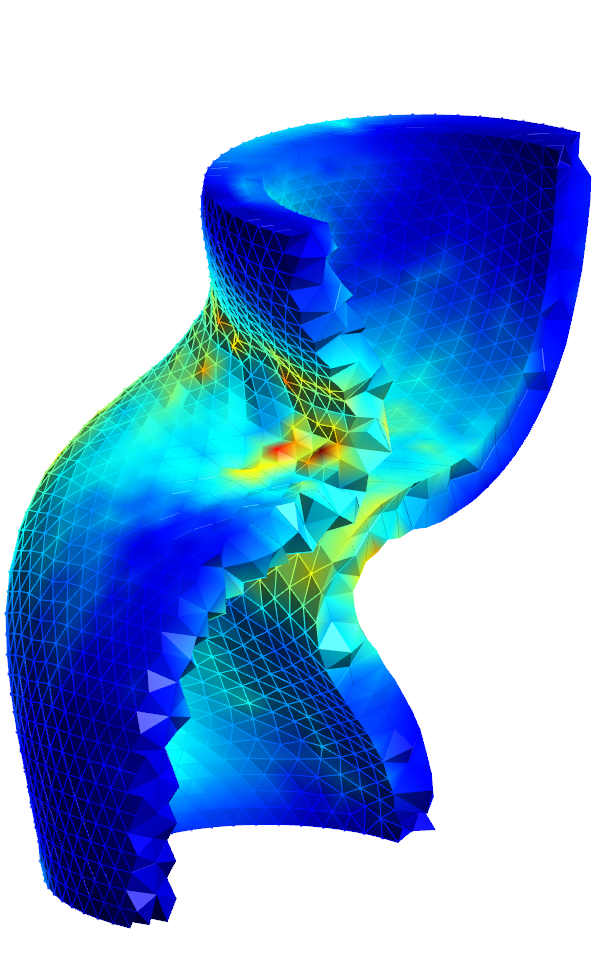} 
    }
    \caption{Sheared cylinder: von Mises stress on the deformed configuration at  $40\%$, $80\%$ and $100\%$ of loading (uHHO(1) solution). The color scales goes from 0.0 (blue) to 0.275 (red).}\label{fig:VM_cylindre}
\end{figure}

\begin{figure}[htbp]
        \centering 
\includegraphics[scale=0.9]{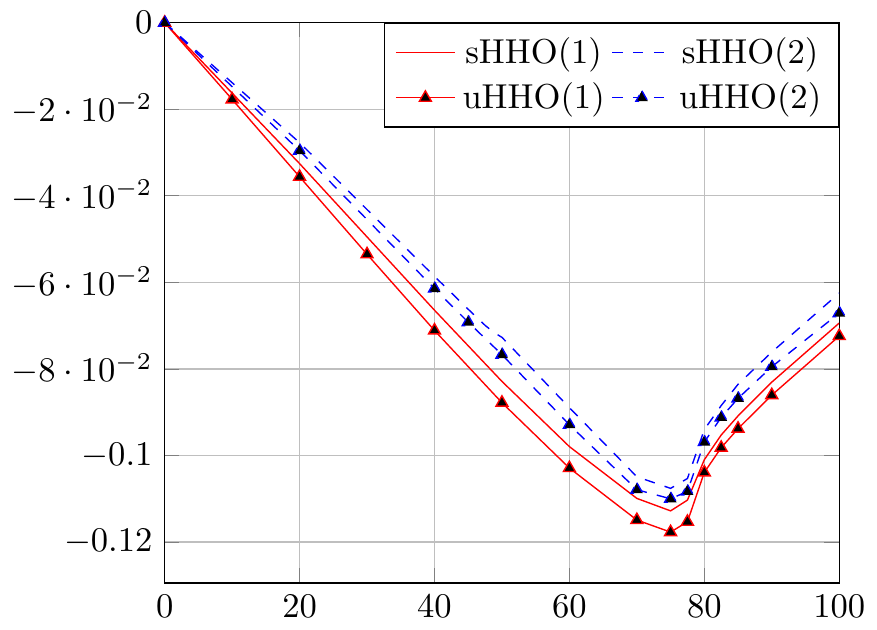}
\caption{Sheared cylinder: evolution during the loading of the vertical component of the discrete traction integrated over the top of the cylinder for sHHO and uHHO.}
  \label{fig::load_cyl}
\end{figure}

This test case, proposed in \cite{KaLeC:2015}, simulates a hollow
cylinder under important compression and shear (it can be seen as a
controlled buckling). The cylinder in its reference configuration has a
inner and outer radius of 0.75 and 1, and a height of 4. The bottom face
is clamped, whereas the top face has an horizontally and vertically
imposed displacement of $-1$ in both directions, and the lateral faces are
traction-free. We set $\mu = 0.1$, $\lambda=1$ (which corresponds to a Poisson ratio of $\nu \simeq 0.455$). For sHHO, the stabilization parameter has to be taken of the order of 
$\beta_0=100$. We notice that both sHHO and uHHO are robust and produce very close
results, which compare very well with the results reported in
\cite{KaLeC:2015}. The loading is applied in 30 steps for uHHO and in
37 steps for sHHO, leading respectively to a total of  152 and 187
Newton's iterations. This indicates that uHHO is up to 20\% more effective for
this test case. 
Some snapshots of the solution obtained with uHHO(1) on a mesh composed of
20382 tetrahedra are shown in Fig.~\ref{fig:def_cylindre} where the color indicates the Euclidean norm
of the displacement. Fig.~\ref{fig:VM_cylindre} displays the von
Mises stress at different loading steps on the deformed configuration. This figure allows one to observe the emerging localization of the deformation field. Finally, the evolution during the loading of the vertical component of the discrete traction integrated over the top face of the cylinder is plotted in Fig.~\ref{fig::load_cyl}. The minimum is reached when the cylinder begins to bend at $75\%$ of the loading; beyond this value, the cylinder becomes less rigid.

\subsection{Sphere with cavitating voids}
The last example simulates the problem of cavitation encountered for
instance in elastomers, that is, the growth of cavities under large
tensile stresses \cite{Ball1982}. Simulations of cavitation
phenomena present difficulties because the growth induces significant
deformations near the cavities. For a review, we refer the reader to
\cite{Xu2011}. Some conforming \cite{Lian2011}, non-conforming
\cite{Xu2011}, and HDG \cite{KaLeC:2015} methods have already been
studied for this problem. For cavitation to take place, the strain
energy density has to be changed, and we consider here, as in \cite{KaLeC:2015},  the following modified Neohookean law:
\begin{equation}\label{CavitaionLaw}
\Psi(\Fdef) = \frac{2\mu}{3^{5/4}}  \left( \Fdef:\Fdef \right)^{3/4} - \mu \ln J + \frac{\lambda}{2} (\ln J)^2,
\end{equation}
where $\mu$ and $\lambda$ are constant parameters. We set $\mu=1$, $\lambda = 1$ (which corresponds to a Poisson ratio of $\nu = 0.25$).

\begin{figure}[htbp]
    \centering
        \includegraphics[scale=0.28]{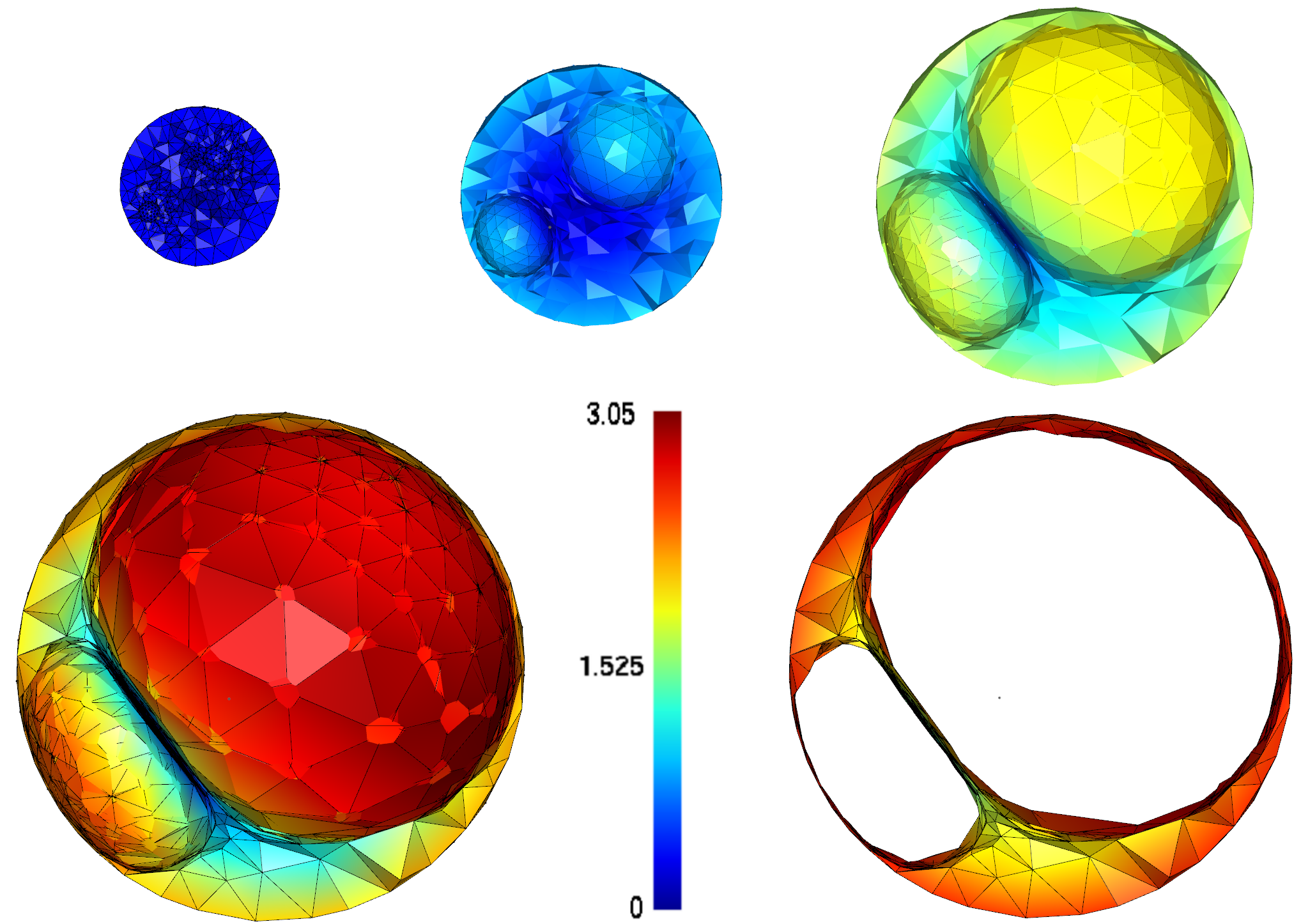}
    \caption{Sphere with cavitating voids: snaphots of the Euclidean displacement norm at $r=0$, $r=0.8$, $r=1.6$ and $r=2.52$ of loading (the sphere is cut along the Equatorial plane) for uHHO(2) on the deformed configuration. The bottom right plot shows a thin slice of the sphere (still along the Equatorial plane) for $r=2.52$.}\label{fig:cavitation3d_def}
\end{figure}

\begin{figure}[htbp]
    \centering
    \subfloat{
        \centering 
\includegraphics[scale=0.9]{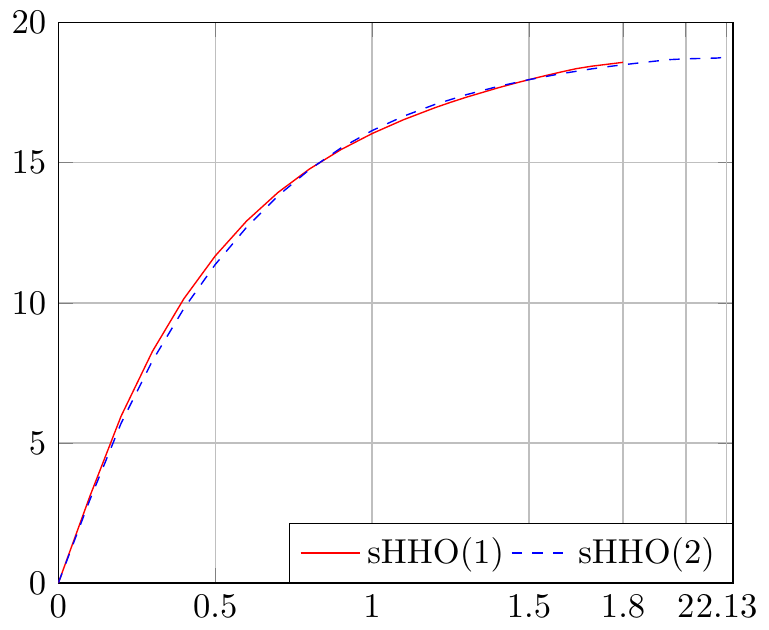}
    }
    ~ 
    \subfloat{
        \centering
\includegraphics[scale=0.9]{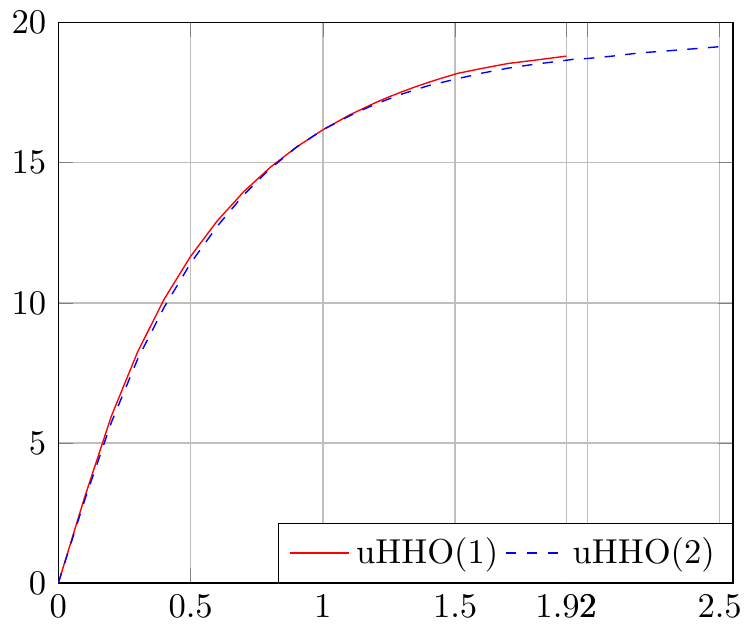}
    }
\caption{Sphere with cavitating voids: radial component of the discrete traction integrated over the outer surface versus the imposed radial displacement obtained with sHHO (left) and uHHO (right). Notice in both cases the larger value attained by $r$ for $k=2$.}
  \label{fig::load_cav}
\end{figure}

The reference configuration consists of a unit sphere of radius 1 with
two spherical cavities. The origin of the Cartesian coordinate system is
the center of the sphere. The first cavity has a radius of 0.15 and its
center is the point of coordinates $(-0.7,-0.7,0)$, and the second
cavity has a radius of 0.2 and its center is the point of coordinates
$(0.25,0.25,0.25)$. A displacement $\vu(\Xo) = r \Xo$ with $r \geq 0$
is imposed on the outer surface ($\abs{\Xo}=1$) of the sphere. The 
stabilization parameter has to be taken of the order of
$\beta_0 = 100$ for sHHO. The mesh is composed of $32288$ tetrahedra,
and the value of $r$ is increased progressively until the moment where the
Newton's method fails to converge. Some snapshots of the Euclidean
displacement norm are shown in Fig.~\ref{fig:cavitation3d_def} on the
deformed configuration for uHHO(2). We also present a zoom near the region where the
two cavities are only separated by a thin layer. The
reported solution compares very well with the HDG solution from
\cite{KaLeC:2015}. Interestingly, the maximum value attained of $r$ is larger for uHHO than for sHHO and is larger for $k=2$ than for $k=1$ (see Fig.~\ref{fig::load_cav}). For $k=2$, the maximum value of $r$ is 2.52 for uHHO and 2.13 for sHHO, which indicates about 15\% more
robustness for uHHO than for sHHO to handle extreme loading situations in this case. Finally, Fig.~\ref{fig::load_cav} presents the radial component of the discrete traction integrated over the outer surface of the sphere versus the imposed radial displacement obtained with sHHO and uHHO.

\section{Conclusion}
We have proposed and evaluated numerically two HHO methods to
approximate hyperelastic materials undergoing finite deformations. Both
methods deliver solutions that compare well to the existing literature on
challenging three-dimensional test cases, such as a hollow cylinder under
compression and shear or a sphere under traction with two cavitating
voids. In addition, both methods remain well-behaved in the
quasi-incompressible limit, as observed numerically on an annulus under
traction and on the indentation of a rectangular block. The test cases
with analytical solution also show that both methods are competitive
with respect to an industrial software using conforming finite elements. The stabilized HHO method rests on a firmer theoretical basis than the unstabilized method, but
requires the introduction and tuning of a stabilization parameter that
can become fairly large in the quasi-incompressible limit. The
unstabilized HHO method avoids any stabilization by introducing a
stable gradient reconstructed in the higher-order polynomial space 
$\Pkpd(T;\Reel^{d\times d})$, but for smooth solutions, the convergence rate is
one order lower than with the stabilized method, i.e., the unstabilized
method still converges, but in a suboptimal way. For
compressible materials, the unstabilized method appears to be somewhat
more competitive than the stabilized method since it requires less Newton's iterations and, at the same time, supports stronger loads, as observed in particular in the case of cavitating voids in the sphere. 

We have also evaluated numerically the unstabilized HHO method using Raviart--Thomas--N\'ed\'elec reconstructions of the gradient (detailed results were not reported herein for brevity). We have retrieved the optimal-order convergence rates for smooth solutions, but the method seems to be somewhat less robust for strongly nonlinear problems. For example, in the case of the sphere with cavitating voids for $k=1$, if the discrete gradient is reconstructed in $\RTNkd(T;\Reel^{d\times d})$, then the maximum value is $r=1.12$ whereas, if the discrete gradient is reconstructed in $\mathbb{RTN}_d^{k+1}(T;\Reel^{d\times d})$ (which contains the space $\Pkpd(T;\Reel^{d\times d})$), then the maximum value for $r$ is the same as for uHHO using $\Pkpd(T;\Reel^{d\times d})$ and $k=1$ ($r=1.92$).

Among possible perspectives of this work, we mention the devising  of a reconstruction based on the ideas introduced in \cite{Eymard2017Preprint} for dG methods, and the use of different reconstructions for the isochoric and volumic parts of the energy density. The present methods can also be applied to approximate other nonlinear problems. For instance, elasto-plasticity constitutes the subject of ongoing work.

\bibliographystyle{abbrv}
\bibliography{Bibliographie}

\end{document}